\documentclass[]{aa}

\pdfoutput=1
\usepackage{graphicx}
\usepackage{txfonts}
\usepackage{amssymb}
\usepackage[font=small]{caption}
\usepackage{indentfirst}
\usepackage{natbib}
\usepackage{booktabs}
\bibpunct{(}{)}{;}{a}{}{,}

\newcommand{\degree}[0]{$^{\circ}$}
\newcommand{\avg}[1]{\left< #1 \right>}         
\newcommand{\cbra}[1]{\left( #1 \right)}      

\newcommand{\hi}{\ion{H}{i}}
\def\Herschel{\textit{Herschel}}
\def\Planck{\textit{Planck}}
\def\IRAS{\textit{IRAS}}

\begin{document}

\title{Structure formation in a colliding flow:\\ The Herschel\thanks{\Herschel\ is an ESA space observatory with science instruments provided by European-led Principal Investigator consortia and with important participation from NASA.} view of the Draco nebula}

\author{
   M.-A. Miville-Desch\^enes\inst{1} \and 
   Q. Salom\'e\inst{2} \and 
   P. G. Martin\inst{3} \and
   G. Joncas\inst{4} \and
   K. Blagrave\inst{3} \and
   K. Dassas\inst{1} \and 
   A. Abergel\inst{1} \and
   A. Beelen\inst{1} \and
   F. Boulanger\inst{1} \and
   G. Lagache\inst{5} \and
   F. J. Lockman\inst{6} \and
   D. J. Marshall\inst{7}
}

\institute{
Institut d'Astrophysique Spatiale, CNRS, Univ. Paris-Sud, Universit\'e Paris-Saclay, B\^at. 121, 91405 Orsay cedex, France \and
   LERMA, Observatoire de Paris, CNRS UMR 8112, 61 avenue de l'Observatoire, 75014 Paris, France \and
Canadian Institute for Theoretical Astrophysics, University of Toronto, 60 St. George Street, Toronto, ON M5S 3H8, Canada \and
D\'epartement de physique, de g\'enie physique et d'optique, Universit\'e Laval, Qu\'ebec, QC G1V 0A6, Canada; Centre de Recherche en Astrophysique du Qu\'ebec (CRAQ), Montr\'eal, QC H3C 3J7, Canada \and
Aix-Marseille Universit\'e, CNRS, LAM (Laboratoire d'Astrophysique de Marseille) UMR 7326, 38 rue Fr\'ed\'eric Joliot-Curie, 13388, Marseille Cedex 13, France \and
National Radio Astronomy Observatory, P.O. Box 2, Rt. 28/92, Green Bank, WV 24944, USA \and
Laboratoire AIM, IRFU/Service d'Astrophysique - CEA/DSM - CNRS - Universit\'e Paris Diderot, B\^at. 709, CEA-Saclay, 91191, Gif-sur-Yvette Cedex, France
}

\titlerunning{Structure formation in a colliding flow: The \Herschel\ view of the Draco nebula}
\authorrunning{Miville-Desch\^enes et al.}

\abstract{The Draco nebula is a high Galactic latitude interstellar cloud observed at velocities corresponding to the intermediate velocity cloud regime. This nebula shows unusually strong CO emission and remarkably high-contrast small-scale structures for such a diffuse high Galactic latitude cloud. The 21 cm emission of the Draco nebula reveals that it is likely to have been formed by the collision of a cloud entering the disk of the Milky Way. Such physical conditions are ideal to study the formation of cold and dense gas in colliding flows of diffuse and warm gas.}
  {The objective of this study is to better understand the process of structure formation in a colliding flow and to describe the effects of matter entering the disk on the interstellar
medium.}
{We conducted \Herschel-SPIRE observations of the Draco nebula. The {\em clumpfind} algorithm was used to identify and characterize the small-scale structures of the cloud.}
{The high-resolution SPIRE map reveals the fragmented structure of the interface between the infalling cloud and the Galactic layer. This front is characterized by a Rayleigh-Taylor (RT) instability structure. From the determination of the typical length of the periodic structure ($2.2$\,pc) we estimated the gas kinematic viscosity. This allowed us to estimate the dissipation scale of the warm neutral medium ($0.1$\,pc), which was found to be compatible with that expected if ambipolar diffusion were the main mechanism of turbulent energy dissipation. 
  The statistical properties of the small-scale structures identified with {\em clumpfind} are found to be typical of that seen in molecular clouds and hydrodynamical turbulence in general. The density of the gas has a log-normal distribution with an average value of $10^3$\,cm$^{-3}$. The typical size of the structures is $0.1$-$0.2$\,pc, but this estimate is limited by the resolution of the observations. The mass of these structures ranges from $0.2$ to $20$\,M$_{\odot}$ and the distribution of the more massive structures follows a power-law $dN/d\log(M) \sim M^{-1.4}$. We identify a mass-size relation with the same exponent as that found in molecular clouds ($M\sim L^{2.3}$). On the other hand, we found that only 15\% of the mass of the cloud is in gravitationally bound structures. }
{We conclude that the collision of diffuse gas from the Galactic halo with the diffuse interstellar medium of the outer layer of the disk is an efficient mechanism for producing dense structures. The increase of pressure induced by the collision is strong enough to trigger the formation of cold neutral medium out of the warm gas. It is likely that ambipolar diffusion is the mechanism dominating the turbulent energy dissipation. In that case the cold structures are a few times larger than the energy dissipation scale. 
The dense structures of Draco are the result of the interplay between magnetohydrodynamical turbulence and thermal instability as self-gravity is not dominating the dynamics. Interestingly they have properties typical of those found in more classical molecular clouds.}


\maketitle


\begin{figure*}[]
  \centering
  \includegraphics[draft=false, width=0.49\linewidth]{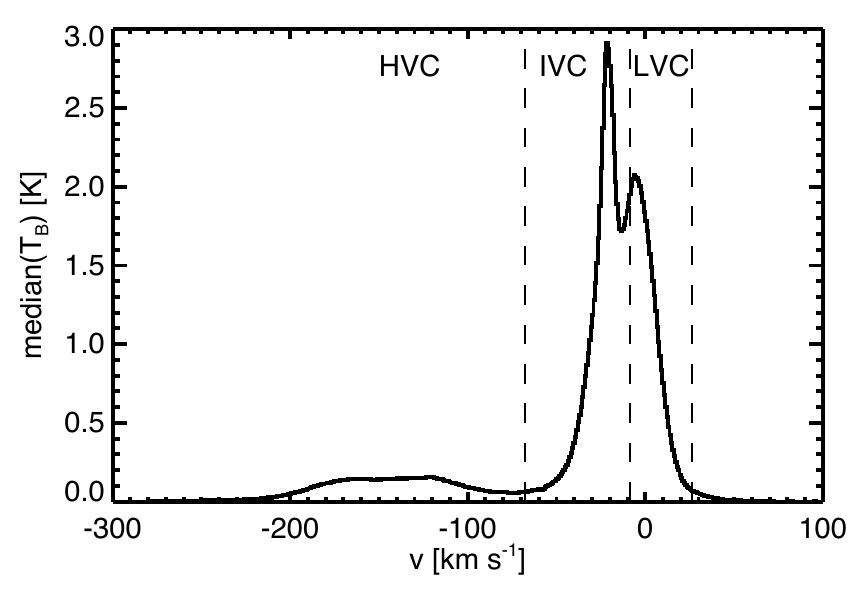}
  \includegraphics[draft=false, width=0.49\linewidth]{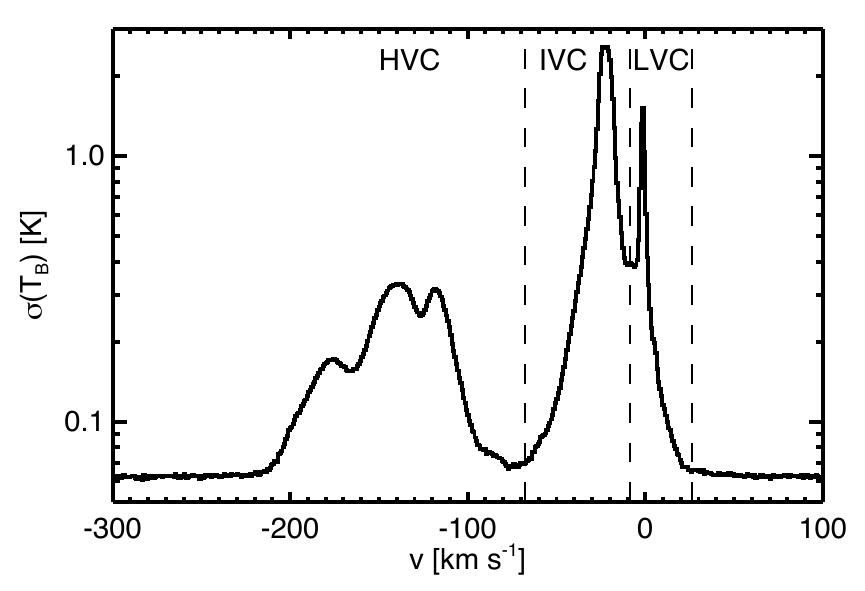}
  \caption{\label{fig:spectrumGBT} Median (left) and standard deviation (right) spectra of the 21\,cm emission in the Draco area. The GBT data used are from \citet{martin2015}. The LVC, IVC, and HVC velocity ranges are indicated by vertical dashed lines.}
\end{figure*}

\section{Introduction}

One of the key questions regarding the formation of stars in galaxies is related to the way gas condenses and how it cycles from the hot and diffuse phase to cold and dense structures where stars form. This process is related to the formation of the cold neutral medium (CNM) and the atomic-molecular (or \hi-H$_2$) transition. One of the frameworks in which this general process is understood is related to colliding flows where a region of the warm neutral medium (WNM) undergoes an increase of pressure that facilitates the rapid cooling of the gas to the stable CNM phase. This process has been studied in several numerical simulations \citep{hennebelle1999,audit2005,vazquez-semadeni2006a,hennebelle2007a,inoue2009,saury2014}. One way of witnessing the formation of cold structures in a colliding flow is by looking at the effects of clouds falling onto the Milky Way disk.

Our Galaxy, as all galaxies, is an open and dynamical system. Matter is constantly arriving on the Galactic disk. Part of this gas is the result of the Galactic fountain: hot gas rises into the halo from stellar winds and supernovae then returns to the disk due to gravity \citep{shapiro1976,bregman1980,putman2012}. Matter also arrives from intergalactic space in the form of either gas stripped from satellite galaxies or gas from the intergalactic medium (IGM). The origin of the material falling into the disk can partly be determined by its metallicity.

The high-velocity clouds (HVC) in our Galaxy are considered possible direct evidence for extragalactic infalling gas. Measurements of the metallicity of HVC gas range from very low (0.1) up to solar \citep{collins2006}. The intermediate-velocity clouds (IVC) are closer to the Galactic layer. Most of these clouds are part of the Galactic fountain, but some should be matter of extragalactic origin that interact with the disk. At this point it is unclear if the amount of new material entering the disk is sufficient to maintain the star formation rate of galaxies \citep{peek2008,sancisi2008}. 

This influx of matter from the halo also adds kinetic energy to the interstellar medium (ISM). It is a way to input the supernovae energy into interstellar turbulence at some distance from where it was produced. One important question relates to the physical properties of the infalling gas and the impact it has on the star formation cycle in galaxies \citep{heitsch2009a,joung2012}. Given the significant velocity (10s to 100s of km\,s$^{-1}$) with which matter encounters the Galactic disk, it is expected to be shocked and undergo dynamical instabilities. The thermal distribution of the gas, its density structure, and the amount of molecular gas produced in the end is not well constrained observationally, partly because of the lack of observations that show this situation clearly.

The Draco nebula (hereafter Draco), an icon among the IVCs, probably provides the best opportunity to understand these processes and resulting physical conditions. By chance there is little structured local ISM gas in the direction of Draco, allowing a very clear view of the matter entering the disk even in integrated (dust or gas) emission. 

In this paper we complement what is already known about Draco by presenting \Herschel-SPIRE \citep{griffin2010a} maps of the nebula, revealing the fine details of its structure, and especially of its Rayleigh-Taylor front. We used these observations to quantify the typical length of the Rayleigh-Taylor instability, putting some constraints on the gas viscosity and properties of turbulence in Draco. The high resolution of the \Herschel-SPIRE data also allows us to characterize the statistical properties of the small-scale structures formed in the postshock region, which enables us to quantify the outcome of the cloud collision in terms of structure formation.

The paper is organized as follows. In Sect.~\ref{sec:draco} we summarize what is known about Draco. Section~\ref{sec:Obs} presents the data used for this study, how we computed the map of the column density, and a description of the structure of the Rayleigh-Taylor instability front.
The small-scale structure is analyzed in Sect.~\ref{sec:structures}. Our results are discussed in Sect.~\ref{sec:discussion} and summarized in Sect.~\ref{sec:conclusion}.

\section{The Draco nebula}

\label{sec:draco}

\begin{figure}[]
  \centering
  \includegraphics[draft=false, width=0.98\linewidth]{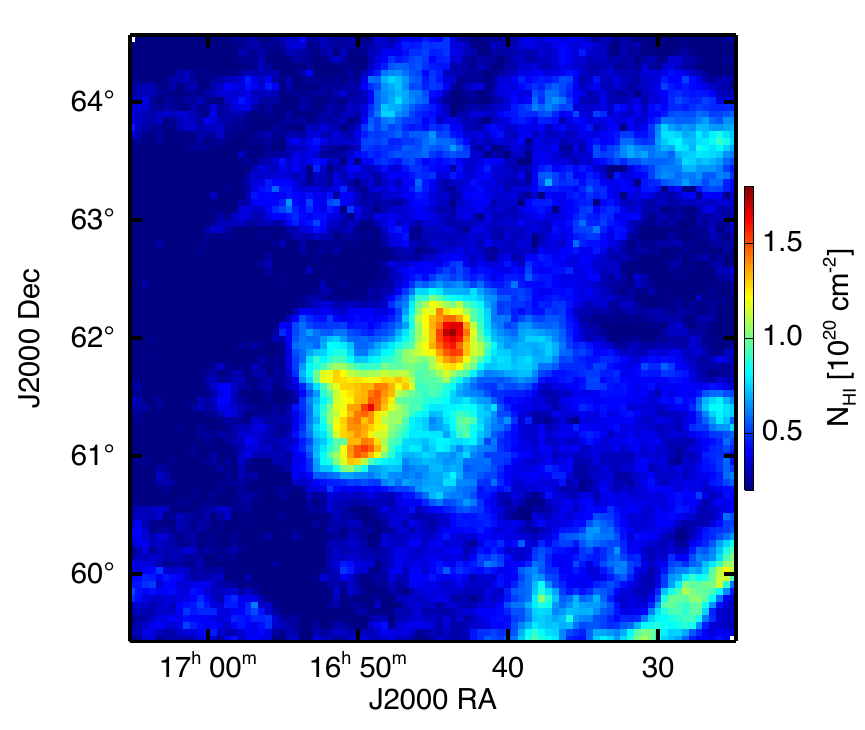}
  \includegraphics[draft=false, width=0.98\linewidth]{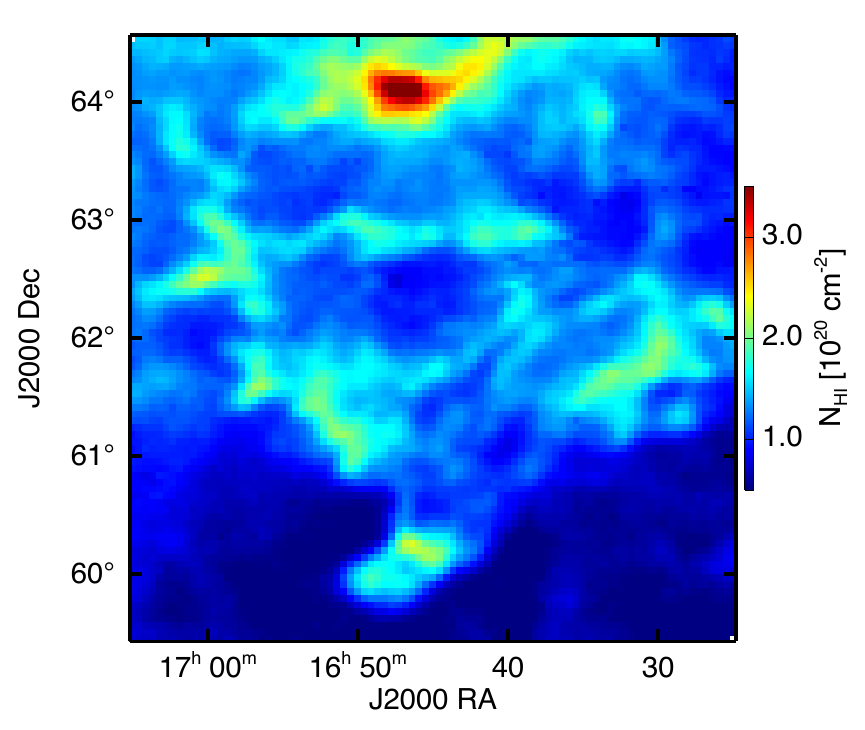}
  \includegraphics[draft=false, width=0.98\linewidth]{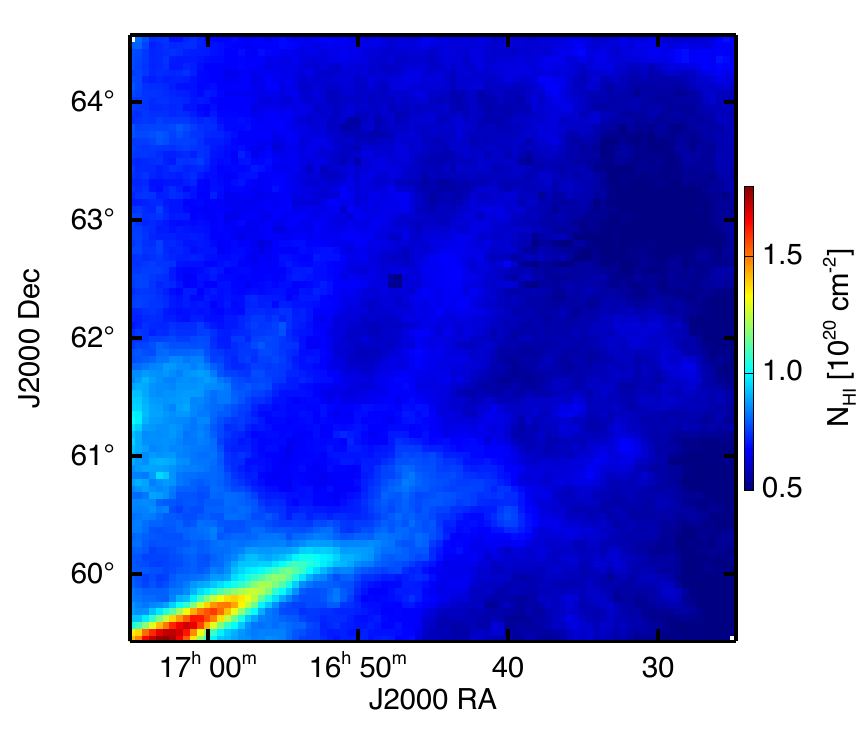}
  \caption{\label{fig:NHI_GBT} \hi\ column density maps in the direction of Draco for the LVC (bottom), IVC (middle), and HVC (top). Data were obtained at the Green Bank Telescope; they are part of the GHIGLS survey \citep{martin2015}. The column density maps are the same as those used in \citet{planck_collaboration2011a}.}
\end{figure}

\subsection{A molecular intermediate velocity cloud}

Draco is the most studied diffuse IVC at high Galactic latitude ($l\approx91$\degree, $b\approx38$\degree). It was first observed at 21\,cm by \cite{goerigk1983} at a velocity $v~\sim-25$\,km\,s$^{-1}$, following calibration observations obtained with the Effelsberg 100\,m telescope \citep{kalberla1982}.
More recently, a $5^\circ\times5^\circ$ region centered on Draco was observed at 21\,cm with the Green Bank Telescope (GBT) as part of the GHIGLS survey \citep{martin2015}. A smaller area was also covered at higher angular resolution with the interferometer of the Dominion Radio Astrophysical Observatory as part of the DHIGLS survey \citep{blagrave2017}. In this region of the sky, three specific \hi\ velocity components are seen: IVC, HVC, and a low-velocity cloud, LVC, which corresponds to the gas in the solar neighborhood. 
These three velocity components show up in the median and standard deviation 21\,cm spectra (Fig.~\ref{fig:spectrumGBT}). Figure~\ref{fig:NHI_GBT} presents their column density map based on the GHIGLS data. To identify the three components, we used the standard deviation spectrum of the 21\,cm emission (Fig.~\ref{fig:spectrumGBT} - right) following \cite{planck_collaboration2011a}.
These maps indicate that the \hi\ column density in this direction of the sky is dominated by Draco, both in absolute value and variations.

\cite{goerigk1983} noticed that the \hi\ cloud coincides with a faint optical nebula seen in the Palomar Observatory Sky Survey (POSS). These authors estimated that the ratio of dust extinction to \hi\ column density, $N_{\rm HI}$, is unusually high, a factor 10 higher than typical values for diffuse clouds, which led them to conclude that most of the gas is in molecular form. From the analysis of their 21\,cm data and its comparison with X-ray measurements, \citet{goerigk1983} also suggested that Draco is the result of the interaction of Galactic halo gas entering the disk and that the details of its front-like structure is the result of the Rayleigh-Taylor (RT) instability. 

The presence of molecular gas was attested by \citet{mebold1985} and \citet{rohlfs1989} who carried CO observations of the brightest part of the nebula and found clumps of strong molecular emission at the boundary of the nebula. CO emission has also been detected by \Planck\ \citep{planck_collaboration2014c}. In addition, \citet{stark1997} reported the detection of CH emission at 3.3\,GHz and \citet{park2009} presented far-UV (FUV) observations showing the molecular hydrogen fluorescence over the whole nebula.

Draco is part of the \citet{magnani1985} catalog of high Galactic latitude molecular clouds (MBM 41 to 44). According to \citet{magnani2010} there are only five other regions at high Galactic latitude where intermediate velocity molecular gas (CO) has been detected; Draco is the region at the lowest Galactic latitude and the largest on the sky (about $4^\circ\times 4^\circ$).

Molecular gas was also revealed indirectly by X-ray data; \citet{moritz1998} estimated that up to 70\% of the hydrogen is molecular in the brightest parts of the nebula. Draco also appears as a shadow in the soft X-ray ROSAT data, enabling to measure the column density of hydrogen independently \citep{burrows1991,snowden1991}. These measurements also imply that most of the gas is in molecular form. Similar conclusions are reached by comparing dust and 21\,cm data \citep{herbstmeier1993,planck_collaboration2011a}. 

One specific feature of Draco is that it shows unusually strong CO emission for a diffuse cloud. It has significant CO emission on lines of sight where the total column density (derived either from dust emission or X-rays) is much lower than the usual threshold of $0.5-1\times 10^{21}\,$cm$^{-2}$ where molecular gas is usually seen in the ISM. In addition, large values of the ratio $W({\rm ^{12}CO})/N({\rm H_2})$ were deduced by \citet{herbstmeier1993,moritz1998}. Also, \citet{herbstmeier1994} observed higher CO transitions (up to J=3-2) to look for unusual excitation conditions but they found line ratios compatible with the average values found in cirrus clouds. This is indicative of a high CO abundance that is potentially caused by an increase of the density due to the collision of a halo cloud entering the Milky Way disk.

\begin{figure*}[]
  \centering
  \includegraphics[draft=false, width=\linewidth]{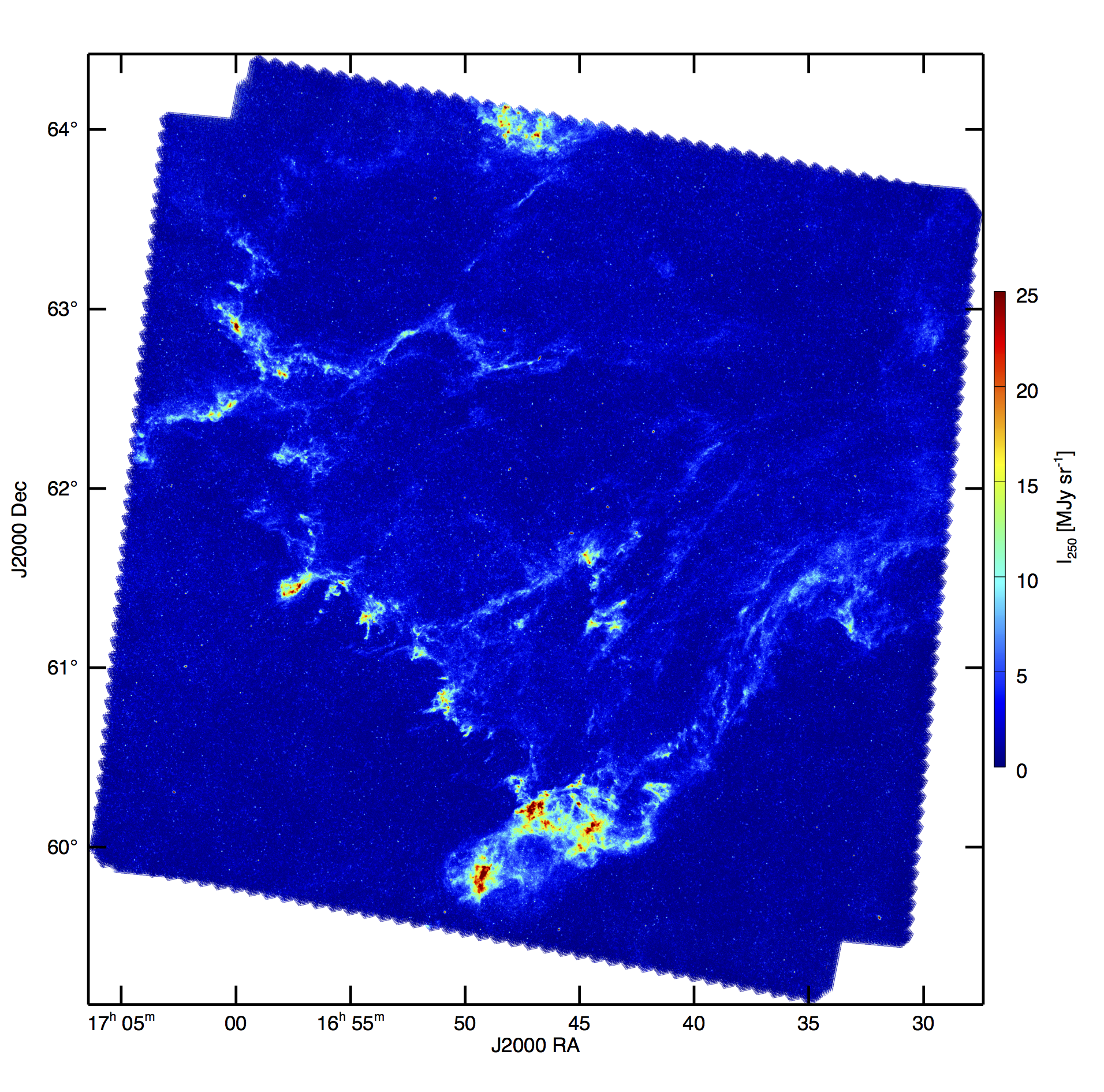}
  \caption{\label{fig:map250} Draco nebula at 250\,$\mu$m as observed with \Herschel-SPIRE.
    From the comparison with \Planck\ and \hi\ data, the 250\,$\mu$m brightness is proportional to
    gas column density (see Section~\ref{sec:Obs}). The color scale shown in this image
    translates to $0 < N_{\rm H} < 6.2 \times 10^{21}$ \, cm$^{-2}$}
\end{figure*}


\subsection{Hot gas}

There have been a few studies looking for hot gas in the area of Draco. \citet{hirth1985} and \citet{kerp1999} showed  an excess of X-ray emission that seems to be spatially correlated with Complex~C, 
but no clear relationship with Draco itself could be established. 
On the other hand, based on FUV observations, \citet{park2009} reported the detection of several ionic lines in the direction of Draco, especially C\,IV, Si\,II, and O\,III]. The C\,IV and Si\,II lines are seen outside Draco, possibly coming from the hot ionized Galactic layer. In addition, there is an excess of C\,IV, correlated with the dust emission of Draco. This excess is also seen in H$\alpha$ and O\,III], but not in Si\,II. Because of the absence of Si\,II, the authors concluded that the C\,IV, H$\alpha,$ and O\,III] emission in Draco cannot be the result of photoionization but it could be from the radiative cooling of warm, shocked gas.

\begin{figure*}[]
  \centering
  \includegraphics[draft=false, width=\linewidth]{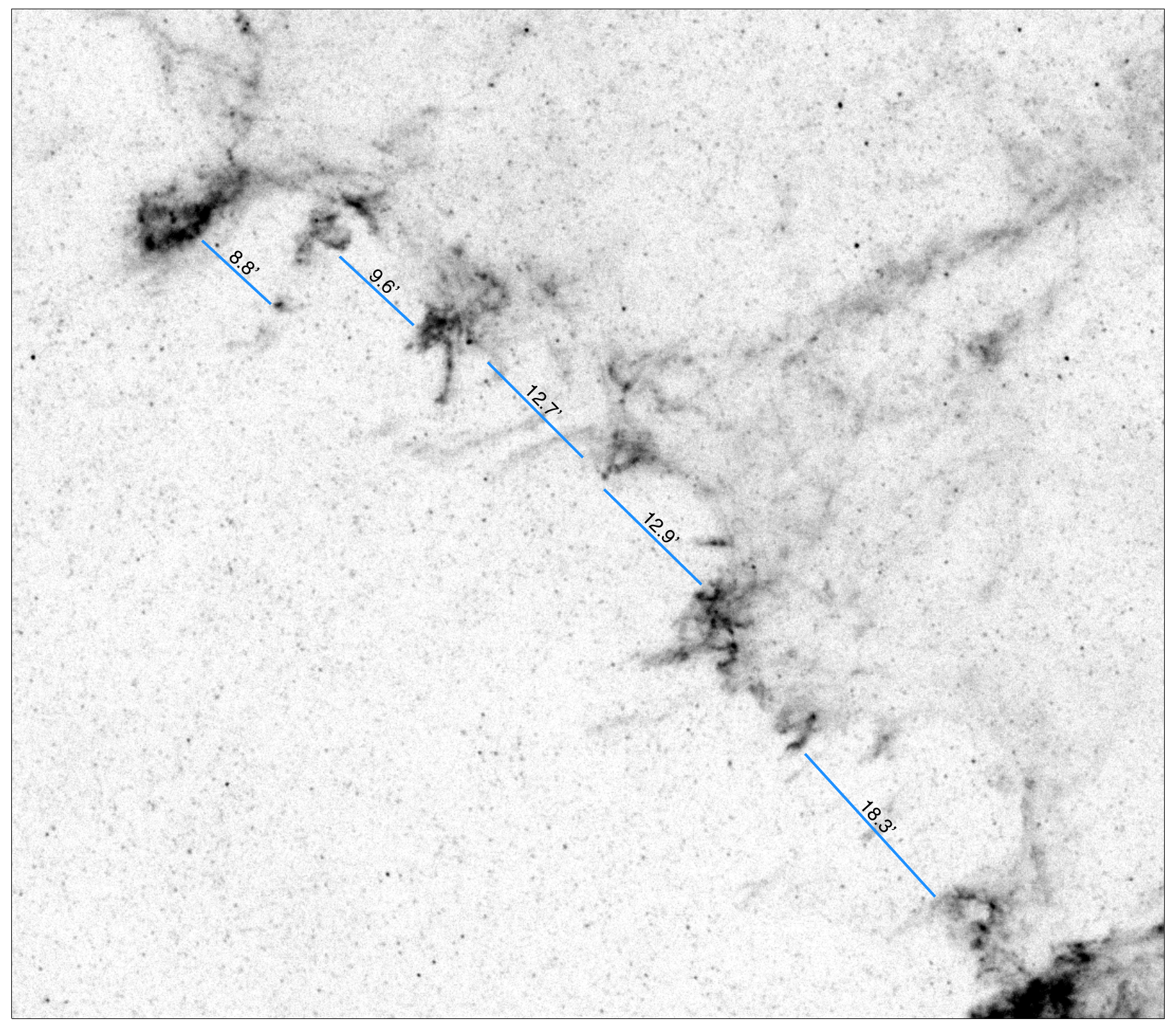}
  \caption{\label{fig:RTI} Rayleigh-Taylor instability structure as seen in the $250$\,$\mu$m SPIRE map. The blue lines give the apparent distance between fingers.}
\end{figure*}

\subsection{Distance}

An important clue informing any self-consistent description is the location of Draco in the Galaxy.
Many attempts to estimate its distance have been made using different techniques.
Star counts and color excess methods were used by \citet{mebold1985,goerigk1986,penprase2000} giving distance estimates of $800 < d < 1300\,$pc. On the other hand, as pointed out by \citet{lilienthal1991}, both these methods have  potential biases especially at low $A_V$. The only direct measure of the distance to Draco came from the detection of Na\,I\,D absorption lines at the systemic velocity of Draco in the spectrum of one star with known parallax distance \citep{gladders1998}. Combined with the nondetection for other foreground stars, \citet{gladders1998} estimated that $463^{+192}_{-136}\le d\le 618^{+243}_{-174}\,$pc.

In the following, we adopt a distance of 600 pc. At the latitude of Draco, this distance corresponds to a height above the Galactic plane of $z= 370\,$pc. As the half width at half maximum of the WNM in the Galactic disk is $\sim 265$\,pc \citep{dickey1990}, Draco is located in the upper part of the diffuse Galactic disk (or lower Galactic halo). What is clear is that Draco is out of the Local Bubble and so it must shadow X-rays coming from the Galactic halo. 

The fact that Draco is located in the outer part of the Galactic disk is compatible with the fact that its brightness in the optical is comparable to clouds with dust properties typical of the disk \citep{mebold1985}. It suggests that Draco is illuminated by a fairly standard interstellar radiation field and so is not likely to be located at kpc distances. A similar argument can be made from its infrared brightness.


\section{\Herschel-SPIRE map of Draco}
\label{sec:Obs}

Draco was observed with \Herschel\ PACS (110 and 170\,$\mu$m) and SPIRE (250, 350 and 500\,$\mu$m) as part of the open-time program {\em ``First steps toward star formation: unveiling the atomic to molecular transition in the diffuse interstellar medium''} (P.I. M-A Miville-Desch\^enes). A field of $3.85^\circ \times 3.85^\circ$ was observed in parallel mode. Unfortunately, an error occurred during the acquisition of the PACS data making them unusable. Therefore, the results presented here are solely based on SPIRE data, especially the 250\,$\mu$m map that has the highest angular resolution.

The SPIRE data were reduced using a standard procedure with HIPE v13. 
We used the product available on the HErSchel IdOc Database (HESIOD\footnote{http://idoc-herschel.ias.u-psud.fr}).
The zero level of each map was set by correlation with \Planck\ data (see Appendix~\ref{sec:convert_to_NH}).
The SPIRE 250\,$\mu$m map of Draco is shown in Fig.~\ref{fig:map250}.
In what follows, we use the 250\,$\mu$m map converted to total hydrogen column density, $N_{\rm H}$. The details of this conversion are given in Appendix~\ref{sec:convert_to_NH}.

  \begin{figure*}[]
  \centering
  \includegraphics[draft=false, width=0.9\linewidth]{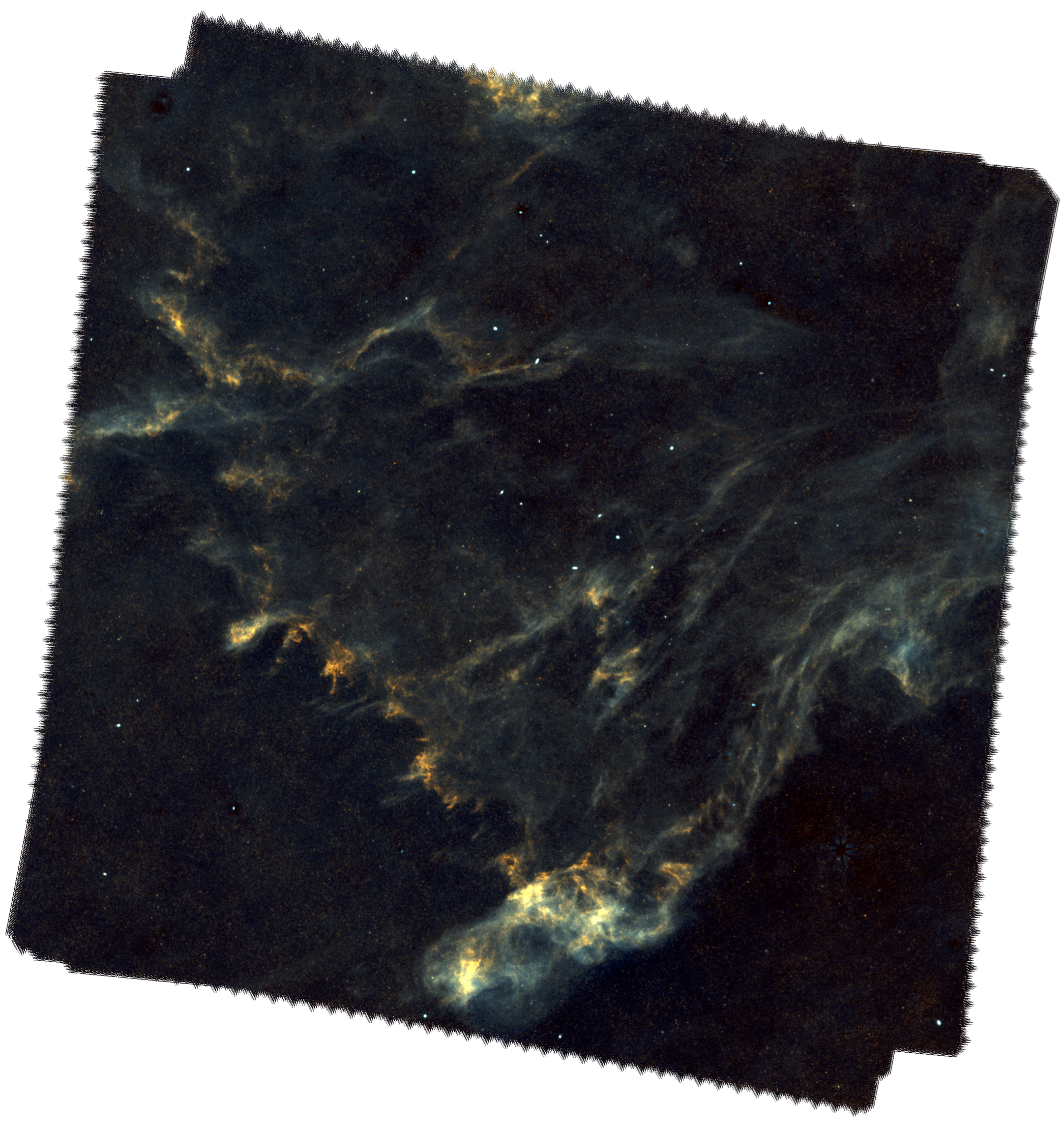}
  \caption{\label{fig:SPIRE_and_WISE} Draco nebula. RGB image combining \Herschel-SPIRE 250\,$\mu$m (red) and WISE 12\,$\mu$m (blue) from \citet{meisner2014}. The green channel is the average of the two others. The size of the mosaic is $\sim 5^\circ \times 5^\circ$. The angular resolution of both maps is comparable (17.6'' for SPIRE 250\,$\mu$m and 15'' for WISE 12\,$\mu$m).}
\end{figure*}

The \Herschel\ data reveals for the first time the structure of matter in Draco at physical scales down to 0.05\,pc (resolution of $17.6''$ at 250\,$\mu$m). The wispy and filamentary structure already seen in previous data (21\,cm, far-infrared) is striking here. At the front structure at low declination where RT type structure were already identified, the SPIRE data reveals a wealth of clumpy structures organized in finger type structures.

One of the most striking features of the \Herschel\ observations of Draco is the structure of the front that shows periodic half shells that are similar to structures produced by the Rayleigh-Taylor instability.
   A close-up view of the shock front is given in Fig. \ref{fig:RTI} revealing the typical arches. The size of the arches is variable and the identification of the fingers is not always obvious. Nevertheless, we found a variation of about a factor of two with an average angular size of $12'.5$. Assuming a distance of 600\,pc and that the structures are orientated perpendicular to the line of sight, the typical length of the instability structure in Draco is $\lambda_{max}\approx 2.2$\,pc. In Sect.~\ref{sec:viscosity} we discuss on how this typical length might provide some information on the viscosity of the gas and on some properties of interstellar turbulence.

The 250\,$\mu$m dust emission map reveals a higher dynamic range of the column density than 21\,cm data. The \hi\ column density (Fig.~\ref{fig:NHI_GBT}) ranges from 1 to $3\times10^{20}$\,cm$^{-2}$, while the dust emission indicates a range from 3 to $50\times10^{20}$\,cm$^{-2}$. It even reaches $N_{\rm H} = 1\times 10^{22}$\,cm$^{-2}$ for a few pixels in the brightest parts in the southern region of the nebula. Part of this difference is due to different angular resolution of the GBT (9') and SPIRE (17.6'') data. Nevertheless, even when brought to the same angular resolution, a significant difference remains between the column density estimated with dust emission and 21\,cm data. This confirms previous studies based on such a comparison and that showed that a significant portion of the bright region of Draco is composed of molecular hydrogen \citep{herbstmeier1993,planck_collaboration2011a}.

Figure~\ref{fig:SPIRE_and_WISE} presents another view that combines SPIRE 250\,$\mu$m and WISE 12~$\mu$m data \citep{meisner2014}. This image highlights strong variations of the relative abundance of smaller dust grains with respect to bigger dust grains. In particular, in the front-like structure, where strong CO emission is observed, there is almost no 12\,$\mu$m emission, indicating a relative lack of smaller dust grains. One possibility is that smaller dust grains have disappeared through coagulation on bigger grains in a denser environment, similar to what is seen in molecular clouds. If that is the case, the color variations in Fig.~\ref{fig:SPIRE_and_WISE} could somehow reflect the variations of the gas density. Other possibilities might relate back to the pre-existing grain populations in the WNM and their response to the collision. The spectacular variations of the dust color ratio from dense to diffuse parts of the nebula encompass rich information on the evolution of interstellar dust in compressed environments. The image shown in Fig.~\ref{fig:SPIRE_and_WISE} clearly deserves a dedicated and detailed analysis that we leave for a future study. 

 \begin{figure}[]
  \centering
  \includegraphics[draft=false]{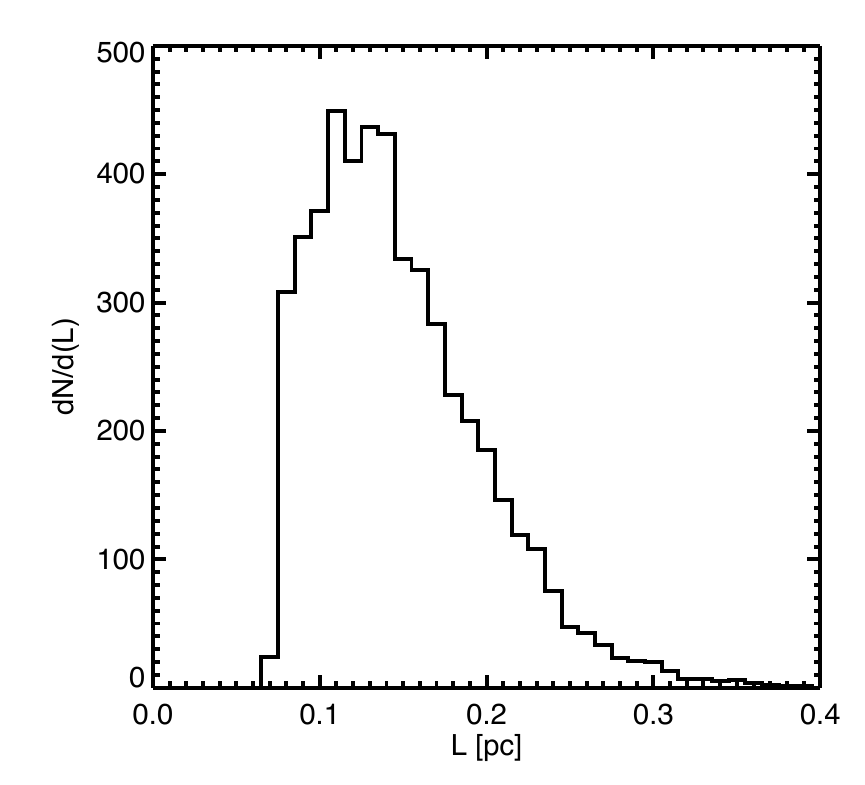}
  \caption{\label{fig:PDF_phys_size} Histogram of the size of each structure computed using the eigenvalues of the inertia matrix.}
\end{figure}

\section{Small-scale structures}
\label{sec:structures}

The \Herschel-SPIRE data provide a striking view of the small-scale structure of Draco 
(Figures~\ref{fig:map250}, \ref{fig:RTI}, and \ref{fig:SPIRE_and_WISE}).
In this section we characterize this structure and compare it with that seen in molecular clouds. 

   \subsection{Structure identification}

   To characterize the general morphology of the nebula we broke it into individual structures using the 2D version of the {\em clumpfind} algorithm \citep{williams1994} that has been used extensively to identify structures in molecular clouds, the Milky Way, and external galaxies. Because it identifies islands above some brightness thresholds, {\em clumpfind} does not make any assumption concerning the shape or size of the structures. On the other hand, as it identifies structures using contours, {\em clumpfind} is rather sensitive to noise because it breaks contours at low brightness level. 

To identify structures on the original 250\,$\mu$m map, we had to set the lowest threshold value of {\em clumpfind} to a value higher than the noise level. The noise level at the pixel size is relatively high compared to the brightness of some of the diffuse emission seen at larger scales. The result was that only the brightest structures were identified by {\em clumpfind}, leaving out more diffuse parts of the nebula that are visible by eye. 
   To minimize the effect of noise, we applied {\em clumpfind} on a version of the map convolved by a Gaussian kernel. Empirically we found that a kernel of FWHM=30.5'', corresponding to a decrease of the angular resolution by a factor of two, from 17.6'' to 35.2'', eliminates the effect of noise. The smoothed map was then projected on a coarser grid with a pixel size that is twice the original size (from 6'' to 12''). 

      We therefore applied {\em clumpfind} on the smoothed \Herschel-SPIRE 250\,$\mu$m map, converted to $N_{\rm H}$, with 40 threshold values, equally spaced in log, from $1.25 \times 10^{21}$ to $2.0 \times 10^{22}$\,cm$^{-2}$ (corresponding to 5 to 80\,MJy\,sr$^{-1}$). The lowest threshold value is rather conservative but it rejects any contamination by galaxies while providing little impact on the properties of the identified structures.
 We stress that the results presented here do not depend significantly on the details of 
these choices (size of the smoothing kernel and threshold levels of {\em clumpfind}). 
   A total of 5028 structures were identified. Their properties are described next.

   \subsection{Physical properties}
   
\begin{figure}[]
  \centering
  \includegraphics[draft=false]{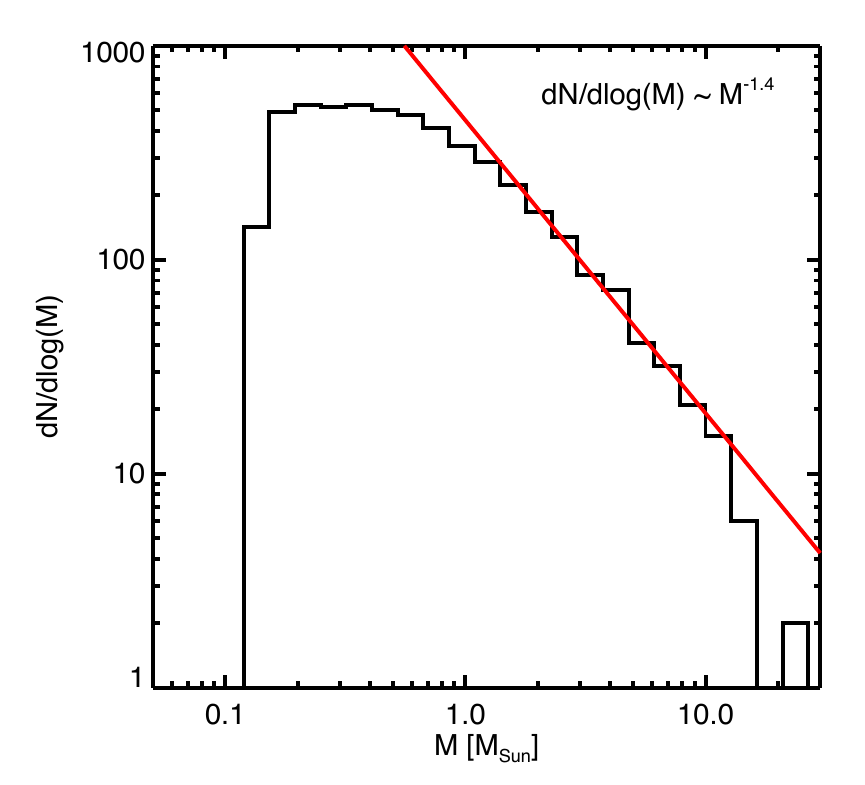}
  \caption{\label{fig:PDF_dN_dlogM} $dN/d \log(M)$ diagram for the identified structures. The red line is a power-law fit for range where $M \geq 2$\,M$_\odot$.}
\end{figure}

\begin{figure}[]
  \centering
  \includegraphics[draft=false]{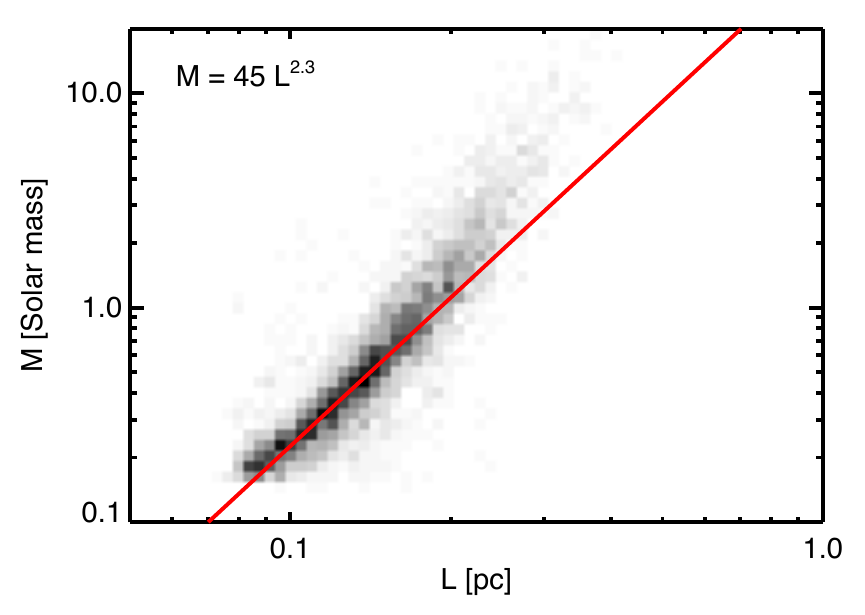}
  \caption{\label{fig:mass_size} Two-dimensional histogram of $M$ vs. $L$. The grayscale is proportional to the density of points. The red line is not a fit. It is indicative of the trend seen at low values of $L$ and $M$.}
\end{figure}

   For each structure we defined a typical size, $L$, its mass, $M$, and its density, $n$.
The histograms of these quantities are shown in Figs.~\ref{fig:PDF_phys_size} to \ref{fig:PDF_density}.

\subsubsection{Size of structures}

For each structure identified we estimated a typical size, taking the brightness distribution of the structure into account; in the definition of the size, brighter pixels have a larger weight. This definition is less sensitive to noise and to the sensitivity of an observation. The method used to estimate $L$ is described in Appendix~\ref{sec:size}.

At a distance of 600\,pc, the angular resolution of the 250\,$\mu$m map translates into a physical distance of 0.05\,pc. This is the FWHM of the smallest structure that can be identified in the map. The way the size is estimated (see Appendix~\ref{sec:size}), the smallest size that can be found corresponds to $L = 0.75\,$FWHM, so the minimum value $L$ can have is in fact 0.04\,pc.
By degrading the resolution of the original image by a factor of two, the minimum $L$ is doubled to 0.08\,pc.
This corresponds exactly to the lowest value of $L$ found here (see Fig.~\ref{fig:PDF_phys_size}). 
The distribution of $L$ has quite a narrow distribution;
50\% of the structures are lower in size than twice the resolution ($L\leq0.16$\,pc)
and 90\% of them are lower in size than three times the resolution ($L\leq 0.24$\,pc).
The largest value of $L$ is 0.47\,pc.

The fact that we are finding physical sizes that are similar to the resolution implies that the emission is varying strongly at small scales. This is compatible with the visual impression given by Figures~\ref{fig:map250}, \ref{fig:SPIRE_and_WISE}, and \ref{fig:RTI}.
The range of sizes found here is likely to be a combination of the angular resolution of the \Herschel-SPIRE data and of how close small-scale structures are. The size of a structure found by {\em clumpfind} is influenced by the distance to its neighbors. The fact that we find that 90\% of the structures are lower in size than $0.25$\,pc is indicative that matter is structured significantly at scales smaller than that value.

   \subsubsection{Mass}
   \label{sec:ClMF}

   The mass of each structure is defined as
   \begin{equation}
     M = N_{\rm Htot} \, D^2 \, \tan^2(\delta)\, \mu m_{\rm H}
   ,\end{equation}
   where $N_{\rm Htot}$ is the total column density of a structure, summed over all pixels,  
   $D$ the distance, $\delta$ the pixel angular size, and $\mu$ the molecular weight. Here we assumed $\mu=1.4$ to take elements heavier than hydrogen into account.

   The total amount of mass in the structures identified is $\sim 5.2\times 10^3\: M_\odot$.
   The mass distribution of the structures is shown in Fig.~\ref{fig:PDF_dN_dlogM}.
   The mass ranges from 0.1 to 20\,$M_\odot$ with a median value of 0.53\,M$_\odot$. 
   We fitted the high mass part of the distribution using a power-law $dN/d\log(M) \propto M^{-\alpha}$ assuming a $1/\sqrt{N}$ uncertainty for each data point.
   The exact value of the power-law exponent depends on the range over which the fit is performed. 
   For $M\geq 2$\,M$_\odot$ we obtain $\alpha=1.4$. 
   We noticed that the low end of the mass distribution has a shape similar to a log-normal distribution.


The value of $\alpha$ found for Draco is significantly different of that found for giant molecular clouds in general \citep[$\alpha \sim 0.8$ for $M>10^4$\,M$_{\odot}$, ][]{solomon1987,kramer1998,heyer2001,marshall2009}. On the other hand, a mass distribution with a similar shape (log-normal plus power-law tail) and a similar range in mass was found in a study of the core mass function (CMF) in Aquila by \citet{konyves2010}. These authors found that the high mass part of their $dN/d\log(M)$ is compatible with a power law with a slope of $\alpha=1.5$. We also note that \citet{peretto2010} found similar $\alpha$ values for small scale fragments of molecular clouds. These similarities are intriguing given the large difference in physical conditions between the cores of Aquila rift, the fragments of molecular clouds, and the clumps of Draco, and considering that different methods were used to identify structures.

We note a general trend between the mass and size of the structures that is similar to what is observed in more massive molecular clouds. Figure~\ref{fig:mass_size} shows a log-log density plot of the mass of the structures versus their size. The low end of the diagram is well modeled by a power-law $M = 45\, L^{2.3}$, where $L$ and $M$ have units of pc and $M_\odot$, respectively.  On the other hand, the structures with sizes larger than 0.2\,pc or masses larger than 1\,$M_\odot$ seem to depart from this relation; these structures are systematically more massive than what the relation predicts.

Several studies have highlighted such a relationship with a very similar exponent. \citet{roman-duval2010} found $M\propto r^{2.26}$ for molecular clouds having sizes in the range 0.7-30\,pc. 
\citet{heithausen1998} found $M\propto r^{2.31}$ for scales ranging from 0.01 to 1\,pc.
A similar mass-size relation is seen in numerical simulations where there is no gravity, and with or without heating and cooling processes included \citep{kritsuk2007,federrath2009,audit2010}. Because of the fact that this relation is seen in different physical conditions, including isothermal gas, it is often attributed to turbulence. 

  \begin{figure}[]
  \centering
  \includegraphics[draft=false]{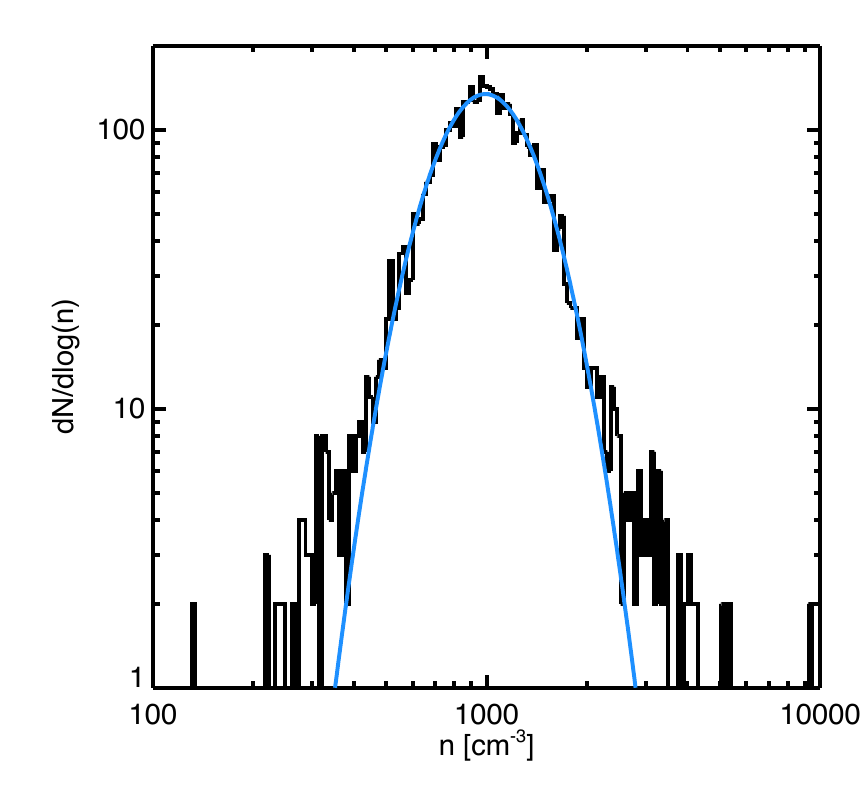}
  \caption{\label{fig:PDF_density} $dN/d\log(n)$ diagram. The blue line is a log-normal fit of the distribution of the density $n$.}
\end{figure}

\subsubsection{Gas density}

The gas density of each structure is defined by
\begin{equation}
n = \frac{3 M}{4\pi L^3} \frac{1}{(\mu+f_{\rm H2}) m_{\rm H}}
,\end{equation}    
where $f_{\rm H_2} = M_{\rm H_2}/( M_{\rm HI} + M_{\rm H_2})$ is the portion of the hydrogen that is in molecular form.  
It is expected that $f_{\rm H2}$ varies across the Draco field; \citet{herbstmeier1993} estimated that the molecular portion varies from 0.04 to 0.7. The details of the \hi-H$_2$ transition in Draco is beyond the scope of the present paper. Here we assume a constant value of $f_{\rm H_2}=0.5$. 

The histogram of $n$ (Fig.~\ref{fig:PDF_density}) is very well represented by a log-normal distribution (i.e., $\log(n)$ is Gaussian distributed)
with a median value of $1.0\times10^3$\,cm$^{-3}$ and a standard deviation of 0.14 in $\log_{10}(n)$. 
The density values given here assume that half of the hydrogen atoms are in H$_2$. 
The majority of the structures have a density higher than the $^{12}$CO $J=1-0$ critical density ($n_{\rm H2} \sim 750$\,cm$^{-3}$). 

In general, it is observed that smaller molecular clouds have higher densities. This is related to the very clumpy and open structure of molecular clouds, which implies a volume filling factor lower than unity. In other words, the mass does not scale with $L^3$ (see Fig.~\ref{fig:mass_size}), therefore the density estimated over large physical scales is an underestimate of the density at small scales, explaining why the average density in GMCs is found to be smaller than the critical density of CO.
For instance, from the sample of molecular clouds of \citet{roman-duval2010}, one obtains that $\avg{n_{\rm H2}} = 800 L^{-0.64}$. For the typical size of a structure in Draco ($L= 0.15$\,pc, see Fig.~\ref{fig:PDF_phys_size}) that relationship would predict $\avg{n_{\rm H2}} = 2.7\times10^3$\,cm$^{-3}$, only a factor of 3 more than what we estimated. Given the large dispersion of the $n_{\rm H_2} - L$ relation observed in the ISM, the values obtained here seem typical for molecular clouds.

   \subsubsection{Jeans mass}

   To evaluate the gravitational stability of the structures in Draco we compared their mass with the 
   Jeans mass that gives the maximal mass of a stable isolated and spherical clouds \citep{lequeux2005},
   \begin{equation}
     M_j = \left(\frac{1}{\mu m_{\rm H}}\right)^2 \, \left( \frac{5}{2} \frac{k T}{G} \right)^{3/2} \,
     \left(\frac{4}{3} \pi n\right)^{-1/2}
     ,\end{equation}
where $\mu$ is the molecular weight (equals to 1.4 or 2.4 for fully atomic or molecular hydrogen, respectively).
   
\begin{figure}[]
     \centering
     \includegraphics[draft=false]{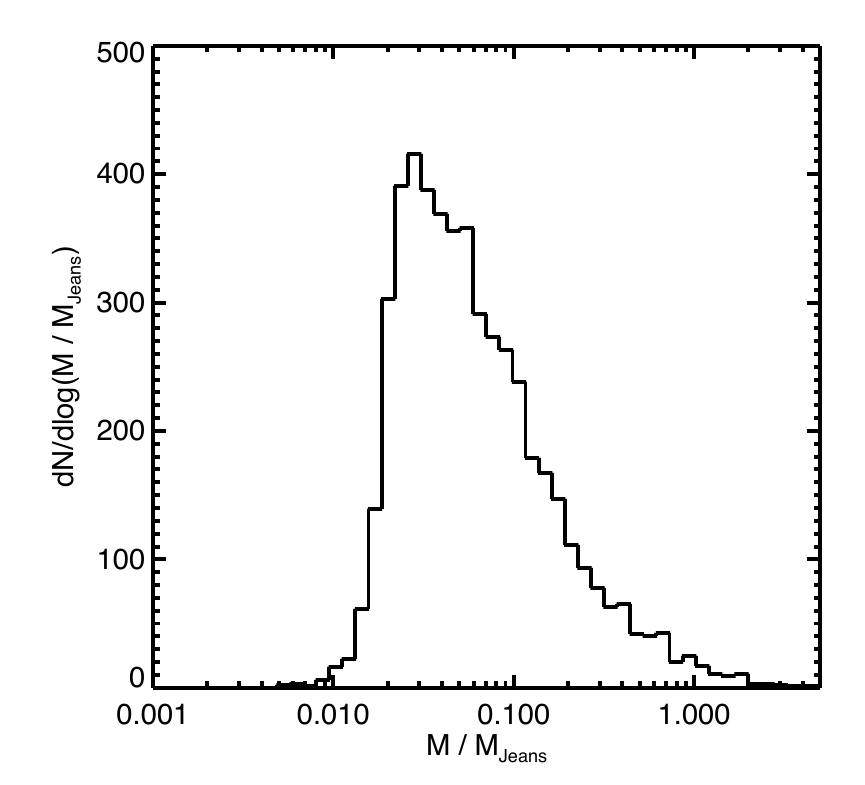}
     \caption{\label{fig:PDF_Jeans_Mass} $dN / d\log(M / M_{\rm Jeans})$ diagram.}
   \end{figure}

   Only 1\% of the structures have a mass larger than the Jeans mass (Fig.~\ref{fig:PDF_Jeans_Mass}). These structures are the more massive. They encompass 15\% of the total mass of the structure identified.

\section{Discussion}
\label{sec:discussion}

The \Herschel-SPIRE observations of Draco reveal two important aspects of the gas structure resulting from the collisions of two flows. The observations reveal a dense front with regularly spaced fingers of dense gas, a structure typical of the RT instability, and numerous and contrasted small-scale structures (clumps). 

\subsection{Formation scenario}

According to what was proposed very early on \citep{kalberla1984,mebold1989,rohlfs1989}, Draco seems to be the result of the compression of warm gas in the outer part of the Galactic WNM layer ($z\sim 370$\,pc) because of the collision with a cloud falling from the Galactic halo onto the disk. Whether the cloud at the origin of the collision was part of the Galactic fountain or from the intergalactic medium is difficult to establish at this point.

\citet{benjamin1997} suggested that IVCs and HVCs have trajectories in the Galactic halo that are compatible with clouds falling on the Galactic disk, attracted by the Milky Way gravitational potential, and slowed down by their interaction with the Galactic ISM. In this scenario, IVCs and HVCs would have reached their terminal velocity.

\begin{figure}
     \begin{center}
     \includegraphics[width=0.98\linewidth, draft=false]{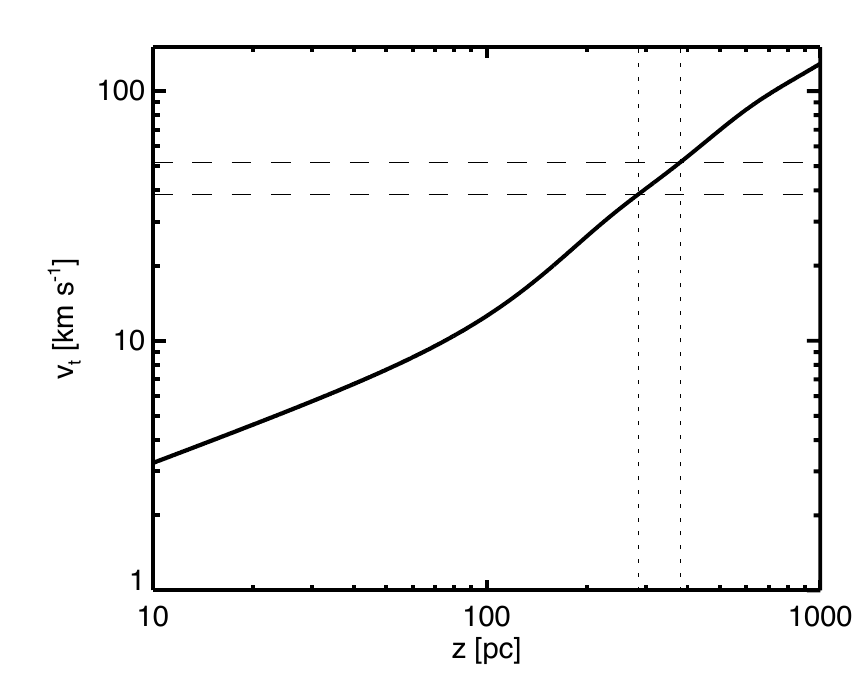}
     \includegraphics[width=0.98\linewidth, draft=false]{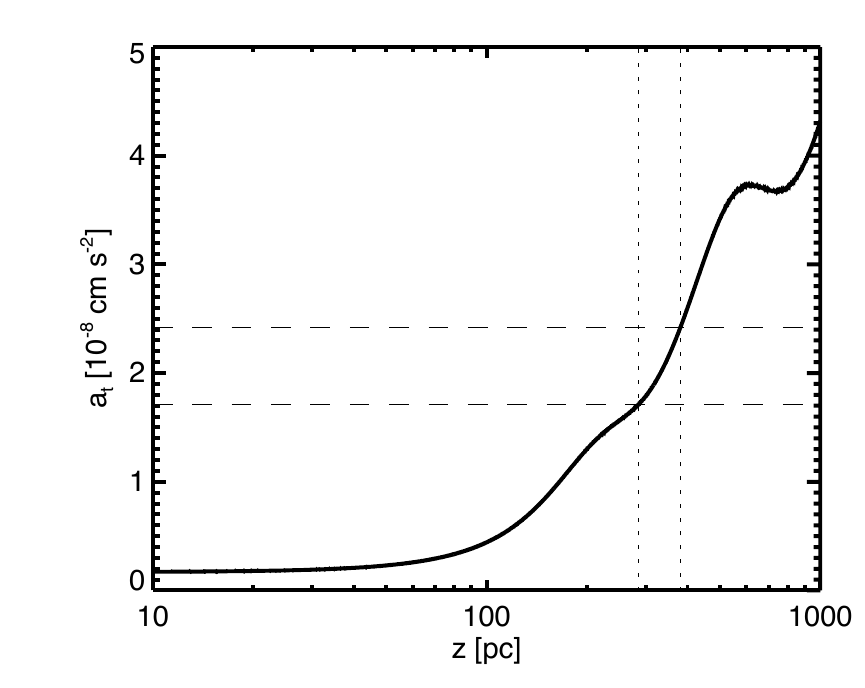}
     \includegraphics[width=0.98\linewidth, draft=false]{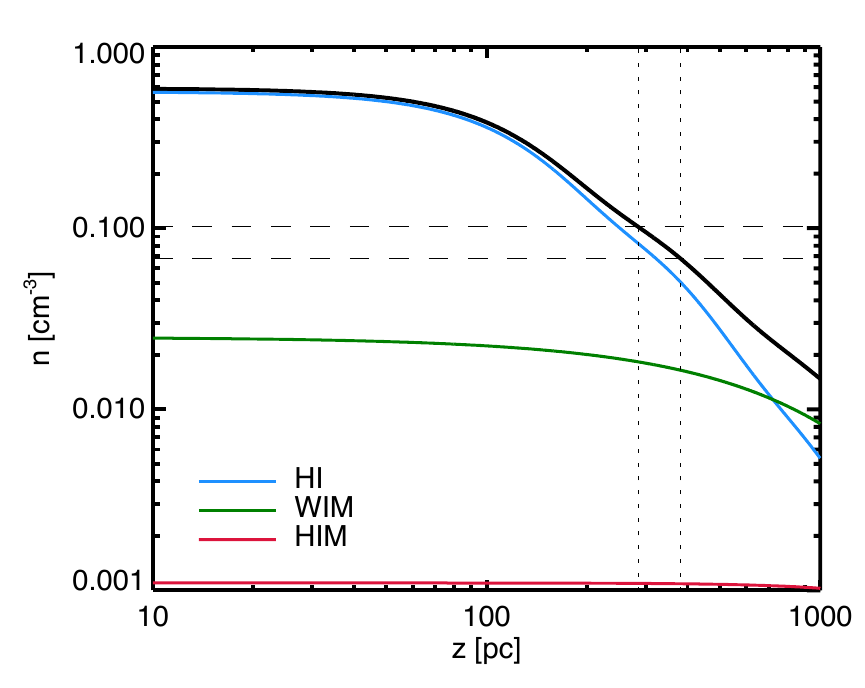}
     \caption{\label{fig:vterminal} Terminal velocity (top) and deceleration (middle) as a function of Galactic height for a cloud of $N_{\rm H}=1.3 \times 10^{20}$\,cm$^{-2}$. The bottom panel is the gas density vs. $z$, including the \hi, the WIM, and the HIM. The dotted lines indicate the range in distance of \citet{gladders1998} and the dashed lines show the corresponding values on each curve.}
     \end{center}
   \end{figure}

According to these authors the terminal velocity depends on the Galactic height, $z$, of the cloud, its column density, $N_{\rm H}$, the gravitational acceleration of the galaxy, $g(z)$,  the density distribution of the halo gas, $n(z)$, and on a parameter that quantify the efficiency of the friction, $C_{\rm d}$,
  \begin{equation}
    \label{eq:terminal_velocity}
v_{\rm t}(z) = \sqrt{\frac{2\, N_{\rm H}\, g(z)}{C_{\rm d}\, n(z)}}.
    \end{equation}
  The total gravitational acceleration (gas and stars) is approximated by
  \begin{equation}
g(z)= 9.5 \times 10^{-9}\, {\rm tanh} (z/{\rm 400 pc}).
  \end{equation}
  The gas distribution is the sum of the \hi\ layer \citep{dickey1990}, the WIM layer \citep{reynolds1993}, and of a hot ionized layer \citep{wolfire1995a}, shown in Fig.~\ref{fig:vterminal} \citep[for details, see][]{benjamin1997}. 
Like in \citet{benjamin1997}, we assumed $C_{\rm d}=1$.

\citet{benjamin1997} stressed that this picture is compatible with the observations as clouds closer to the plane have lower absolute velocity than clouds far away. Draco fits that picture. Assuming a column density of $N_{\rm H} = 1.3 \times 10^{20}$\,cm$^{-2}$ and a height above the plane of $z = 370$\,pc, its observed velocity ($v= v_{\rm LSR}/\sin b = 40$\,km\,s$^{-1}$) is comparable to its expected terminal velocity \citep[$v_{\rm t}\approx 50$\,km\,s$^{-1}$, see Fig.~\ref{fig:vterminal} and ][]{benjamin1999a}.

If Draco has indeed reached its terminal velocity, its average velocity does not inform us on its origin (Galactic fountain or extragalactic gas). Draco could have once been an HVC but it would have slowed down as it interacted with the ISM of the disk. One fact that could favor the Galactic fountain origin is the bright dust and CO emission, showing that it contains a significant amount of heavy elements. On the other hand, the matter that composes Draco today could be Galactic gas swept up by an originally low metallicity cloud that has been slowed down by the Galactic ISM.
At $z=370$\,pc above the Galactic plane, the total column density (\hi+H$^+$) encountered by a cloud coming from infinity is about $N_{\rm H} = 1\times10^{20}$\,cm$^{-2}$, estimated by integrating $n(z)$ (Fig.~\ref{fig:vterminal}) for $z>370$\,pc. This is similar to the column density of Draco itself.
Therefore, a large portion of the mass of Draco could be composed of Galactic gas with near solar metallicity even though the original cloud had a low metallicity.
A detailed study of the metallicity of the gas in Draco could provide some answers to that question.

There has been some speculation about the potential role of the HVC component seen at $v_{\rm LSR} \sim -150$\,km\,s$^{-1}$ in the formation of Draco \citep{hirth1985}. The morphology of the three \hi\ components (Fig.~\ref{fig:NHI_GBT}) could suggest that the IVC (Draco) is the result of the dynamical interaction between the HVC and LVC. 
Along those lines, \citet{pietz1996} proposed that faint 21\,cm emission at velocities between the HVC and IVC could be due to an inelastic collision of HVC gas with the Galactic thick disk. This emission is extremely faint (see Fig.~\ref{fig:spectrumGBT}). 
If Draco is the result of gas being decelerated from -100 to -20\,km\,s$^{-1}$, one would expect to see significant 21\,cm emission at all velocities. One possibility could be that 
the original HVC was composed of two components: one component that has already hit the disk and produced Draco and a second component that is on its way and that we observe today at $v<-100$\,km\,s$^{-1}$.
 \\

Whatever the origin and metallicity of the infalling cloud, it seems likely that it has
accelerated and pressurized WNM gas at rest in the Galactic layer, producing a shocked front of denser gas that is progressing toward us in the diffuse ISM. The structure of Draco itself, with its long cometary plumes, is indeed reminiscent of a dense cloud moving in a more diffuse medium \citep{odenwald1987}.
The structure of the front, with its sharp increase in column density, is also indicative of a shock. 
Assuming $T=8000$\,K and taking elements heavier than hydrogen into account, Draco moves with respect to the Galactic ISM with a Mach number of $M=5.8$. This scenario is compatible with the FUV observations of \citet{park2009} that show the presence of hot gas due to shock heating. In addition, these conditions are favorable for the development of a RT instability at the interface and for the formation of CNM out of compressed WNM through the thermal instability \citep{saury2014}.

\subsection{Turbulence}

In this section we use the RT typical length, combined with velocity information from the 21\,cm data, to estimate parameters of interstellar turbulence, such as the rate of transferred energy, $\epsilon$, the Reynolds number, $Re$, and the turbulence dissipation scale, $l_d$.
The values of these parameters estimated for Draco are summarized in Table~\ref{tab:turbulence}.

   \subsubsection{Rayleigh-Taylor instability and gas viscosity}

   \label{sec:viscosity}

The RT instability occurs when two fluids are accelerated toward each other. 
For the incompressible case, the fluids cannot interpenetrate freely; bubbles of the light fluid rise into the heavy fluid forming finger structures. It appears that such structures are also observed in compressible fluids, like supernovae remnants \citep{ellinger2012}. In our case, the fingers are pointing toward the Galactic plane, indicating that Draco is denser than the medium in which it is moving toward, i.e., the Galactic WNM layer.

   For two incompressible fluids of constant viscosity separated by a horizontal boundary, \cite{chandrasekhar1961} showed that the RT instability typical length is given by
\begin{equation} \label{eq:Chandrasekhar}
  \lambda_{\rm max}=4\pi \cbra{\frac{\nu_{\rm kin}^2A}{a}}^{1/3}
,\end{equation}
where $\nu_{\rm kin}$ is the kinematic viscosity, $a$ is the acceleration, and
\begin{equation}
  A=\frac{\rho_1-\rho_2}{\rho_1+\rho_2}
  \end{equation}
is the Atwood number that takes values between 0 and 1.
Here $\rho_1$ and $\rho_2$ are the density of the heavier and lighter fluids, respectively.

   The Chandrasekhar study of hydrodynamic stability was made for incompressible fluids, which is not the case in the interstellar medium. However, \cite{ribeyre2004} studied the compressible case for supernovae and concluded that compressibility slows down the growth of the RT instability, but it has no important impact on $\lambda_{\rm max}$. The typical length of the RT instability in supernovae is usually estimated with equation \ref{eq:Chandrasekhar} \citep{ellinger2012}. The same assumption is made here.
Interestingly, Eq.~\ref{eq:Chandrasekhar} offers us the opportunity to estimate the gas viscosity knowing the instability typical length, the gas acceleration, and $A$.

What acceleration should be considered here ? In the classical picture of two fluids, one on top of the other, $a$ is the gravitational acceleration. 
Here the gas is slowed down by the friction with the Galactic ISM. In fact the cloud is decelerating as it gets closer to the disk, slowly reaching $v=0$.

  In this context, a cloud entering the disk is slowed down by the friction with the ISM.
  The deceleration can be estimated by
  \begin{equation}
a_{\rm t} = v_{\rm t} \, \frac{d v_{\rm t}}{dz}.
    \end{equation}
Assuming that Draco has reached its terminal velocity, and using Eq.~\ref{eq:terminal_velocity} for $v_{\rm t}$, we estimated that its deceleration is $a=2.3\times 10^{-8}$\,cm\,s$^{-2}$.

The parameter $A$ depends on the density of the two fluids. The density of the gas at a height above the plane of $z\approx370$\,pc, the assumed location of Draco, is $n=0.07$\,cm$^{-3}$ (see Fig.~\ref{fig:vterminal}). For Draco, the density can be estimated using the dust emission and 21\,cm data. The gas column density, averaged over the whole nebula, is about $1.3 \times 10^{20}$\,cm$^{-2}$. Draco spans about 3$^\circ$ on the sky, which translates into 30\,pc at a distance of 600\,pc. Assuming a depth of 30\,pc for the infalling cloud, its average density would be on the order of 1.4\,cm$^{-3}$. This is an averaged value that represents the density of the WNM in Draco. 
The density is certainly larger at the shock front where strong CO emission \citep[as well as CH,][]{stark1997} is observed (the CO critical density is on the order of $10^3$\,cm$^{-3}$).
Assuming $n_1=1.4$\,cm$^{-3}$ and $n_2=0.07$\,cm$^{-3}$, the Atwood number is $A=0.90$. 

The RT length, $\lambda_{\rm max}$, is the one estimated from the periodic structure observed at the dense front (Fig.~\ref{fig:RTI}). Even though describing the front as RT-like structure is appealing, the determination of the typical scale is difficult and somewhat uncertain. The arch-like structures have sizes ranging from 1.5 to 3.2\,pc, with an average value of $\lambda_{\rm max}=2.2$. Using this average value, the kinematic viscosity computed with Eq.~\ref{eq:Chandrasekhar} is $\nu_{\rm kin} = 6.3 \times 10^{22}$\,cm$^2$\,s$^{-1}$.

It is interesting to compare $\nu_{\rm kin}$ to the more classical molecular viscosity
   \begin{equation}
     \nu_{\rm mol}=\frac{1}{3}\frac{1}{\sigma n}\sqrt{\frac{3}{2}\frac{kT}{\mu m_{\rm H}}}.
   \end{equation}
   For typical values of temperature and density for the CNM ($T=100$\,K, $n=30$\,cm$^{-3}$) and WNM in Draco ($T=8000$\,K, $n=1.4$\,cm$^{-3}$) and assuming $\sigma=1\times 10^{-15}$\,cm$^{-2}$ for the hydrogren cross-section \citep{lequeux2005}, the molecular viscosity is $\nu_{\rm mol}=1.0 \times 10^{18}$\,cm$^2$\,s$^{-1}$ and $\nu_{\rm mol} = 2.0\times 10^{20}$\,cm$^2$\,s$^{-1}$ for the CNM and WNM, respectively.
   The value of the WNM (the largest of the two) is more than 300 times smaller than what is estimated using the RT typical length and Eq.~\ref{eq:Chandrasekhar}. 
   That results depends on the distance to the power $3/2$. Even if we assumed a distance to Draco of 400\,pc, the kinematic viscosity would still be more than 200 times larger than the molecular viscosity. {This is also much larger than the uncertainty of less than a factor of 2 on $\lambda_{\rm max}$.}

\subsubsection{Energy transfer rate}

An important parameter that characterizes the turbulent cascade of energy is the energy transfer rate by unit of mass,
   \begin{equation}
\epsilon = \frac{1}{2} \frac{\bar{v_l}^3}{l}.
   \end{equation}
Here $l$ and $\bar{v_l}$ represent the typical scale and typical average velocity of the flow.  We assumed $l \sim 30$\,pc for the largest scale of the flow. The average velocity of the gas at that scale is best estimated by looking at the width of the 21\,cm line, averaged over the whole region. Using the 21\,cm GBT data (Fig.~\ref{fig:NHI_GBT}), we estimated the velocity dispersion of the IVC component to be $\sigma_v=10.7$\,km\,s$^{-1}$, which is likely to be dominated by the warm phase. Removing the thermal contribution to the line width (assuming $T=8000$\,K), this reduces to $\sigma_v=7.0$\,km\,s$^{-1}$. Assuming a Gaussian distribution of velocity, this corresponds to an average velocity of $\bar{v_l} = \sqrt{8/\pi}\,\sigma_v=11.2$\,km\,s$^{-1}$. 

The energy transfer rate of the WNM in Draco is $\epsilon = 3.9\times 10^{-3}$\,$L_\odot$\,$M_\odot^{-1}$. This value is slightly higher than that estimated for the \hi\ in the solar neighborhood \citep[$\epsilon \sim 10^{-3}$\,$L_\odot$\,$M_\odot^{-1}$;][]{hennebelle2012a}. 

\subsubsection{Reynolds number}

By combining velocity information from the 21\,cm data with the measurement of $\nu_{\rm kin}$, one can estimate the Reynolds number of the flow
\begin{equation}
Re = \frac{l\, \bar{v_l}}{\nu_{\rm kin}}.
\end{equation}
For Draco, we find $Re=1600$, a value significantly smaller than the more traditional estimates for the ISM \citep[$Re \sim 10^5$;][]{chandrasekhar1949}.
This difference is explained by the fact that $\nu_{\rm kin} >> \nu_{\rm mol}$ as
$Re$ is usually computed assuming that the dissipation of turbulent energy is dominated by molecular viscosity.

\begin{table}
  \caption{\label{tab:turbulence} Properties of the turbulence in Draco.}
\tabskip=0pt
  \begin{center}
    \begin{tabular}{llll}\specialrule{\lightrulewidth}{0pt}{0pt} \specialrule{\lightrulewidth}{1.5pt}{\belowrulesep}
    Quantity & Symbol & Value & Units \\  \midrule
    Largest scale & $L$ & 30 & pc\\
    Average velocity & $\bar{v_l}$ & 11.2 & km\,s$^{-1}$\\
    Rayleigh-Taylor length & $\lambda_{\rm max}$ & 2.2 & pc\\
    Viscosity & $\nu_{\rm kin}$ & $6.3\times10^{22}$ & cm$^2$\,s$^{-1}$\\
    Reynolds number & Re & $1600$ \\
    Energy transfer rate & $\epsilon$ & $3.9\times10^{-3}$ & $L_\odot$\,$M_\odot^{-1}$ \\
    Dissipation scale & $l_d$ & $0.1$ & pc \\ \bottomrule[\lightrulewidth]
  \end{tabular}
  \end{center}
  \end{table}

\subsubsection{Dissipation scale}

Turbulent energy in the ISM is dissipated via either viscous (hydrodynamical) or resistive (MHD) processes, depending on which process happens at the largest scale \citep{benjamin1999a,hennebelle2013c}.
The scale at which turbulent energy is dissipated is 
\begin{equation}
  l_{\rm d} \approx \left( \frac{\nu^3}{\epsilon} \right)^{1/4}.
  \end{equation}
Like for $Re$, the scale at which turbulent energy is dissipated depends on what process is more efficient at transforming motion into heat. 
Assuming that $\nu=\nu_{\rm kin}$, the dissipation scale is $l_d=0.1$\,pc. 
This is close to two to three orders of magnitude larger than what is generally estimated for the WNM ($l_d \sim 0.003$\,pc) and the CNM ($l_d \sim 2$\,AU), assuming that the dissipative process is molecular viscosity \citep{lequeux2005}. 

The other potential dissipative process is the ion-neutral friction.
Following \citet{lequeux2005}, the ambipolar diffusion typical scale is 
\begin{equation}
  \label{eq:ldiss}
l_{\rm AD} = \sqrt{\frac{\pi}{\mu m_{\rm H}}} \frac{B}{2X\,\langle\sigma v\rangle \, n_n^{3/2}}
  ,\end{equation}
where $X$ is the ionization ratio, $n_n$ the density of neutrals, and $\langle \sigma v \rangle$ the collision rate between ions and neutral (assumed to be the Langevin rate $2 \times 10^{-9}$\,cm$^3$\,s$^{-1}$).
For typical conditions in the ISM, from warm gas to molecular clouds, $l_{\rm AD}$ is often larger than the dissipation scale for molecular viscosity. This is the case in Draco; assuming $n_n=1.4$\,cm$^{-3}$ (the density of the compressed WNM) and $B=6$\,$\mu$G, which is a typical value for the WNM in the solar neighborhood \citep{beck2001}, the ambipolar diffusion scale is also $l_{\rm AD}=0.1$\,pc.

The kinematic viscosity estimated from the RT-like structures indicates that the dissipation of energy happens at a scale of $\sim 0.1$\,pc; this is much larger than the dissipation scale of molecular viscosity but is in agreement with what is expected if ambipolar diffusion is the dominant process of energy dissipation. 

\subsubsection{Wardle instability}

Given that the magnetic field seems to play an important role in dynamics, one could consider that the structure of the front is the result of the Wardle instability \citep{wardle1990}. This instability occurs because of the velocity difference between neutrals and ions, when the Alfven Mach number, $M_A$, is greater than 5. The Alfven speed in a WNM with $n=0.1$\,cm$^{-3}$ and $B=6$\,$\mu$G is $V_A = 35.0$\,km\,s$^{-1}$. Considering that the velocity of the shock is 40\,km\,s$^{-1}$, the Alfven Mach number is $M_A=1.1$. This suggests that the Wardle instability is not dominant in Draco. 

\subsection{Dense structures}

Using the two-dimensional version of {\em clumpfind} on the \Herschel-SPIRE 250\,$\mu$m, converted to $N_{\rm H}$, structures of sizes from 0.08 to 0.5\,pc were revealed. 
Interestingly, these small structures share many properties with sub-structures found in more massive molecular clouds as well as with prestellar cores. 
First, the mass distribution ranging from 0.1 to 20\,$M_\odot$ with a median value of 0.53\,$M_\odot$ is very similar to a prestellar core mass distribution \citep{konyves2010}. In addition, the mass-size relation ($M\propto L^{2.3}$)
is typical of that seen in more massive molecular clouds. Figure~\ref{fig:SPIRE_and_WISE} also shows that the 12\,$\mu$m emission from smaller dust grains is extremely weak compared to the big grain emisson at 250\,$\mu$m, as is observed in dense parts of molecular clouds \citep{stepnik2003}.

On the other hand, unlike prestellar cores \citep{konyves2010} and unlike molecular clouds \citep{roman-duval2010}, 85\% of the structures found in Draco are not gravitationally bound. This is compatible with the fact that the density histogram is found to be log-normal, something generally associated with compressive supersonic flows without gravity \citep{kritsuk2007}.
   Also, even though the structures in Draco are dense enough ($n\sim 1000$\,cm$^{-3}$) to explain the strong CO emission observed, the density of these structures is 100 times less than that seen in prestellar cores.

We suggest that the formation of dense structures seen in Draco is the result of the transition from the WNM to cold and dense gas through the thermal instability of \hi. This transition is favored by the compression of the warm gas at the interface between the Galactic layer and the infalling matter. As shown in several numerical simulations, the increase of density of the WNM sends the gas in a thermally unstable state that favors the transition to the CNM \citep{hennebelle1999,audit2005,vazquez-semadeni2006a,inoue2009,saury2014}.
If the increase in density is sufficient, the formation of molecules is expected; using hydrodynamic simulations of a strong shock wave propagating into WNM, \cite{koyama2000} showed that it can produce a thin and dense H$_2$ layer via thermal instability. They also predicted fragmentation of the thermally collapsed layer into small molecular structures, similar to that seen in Draco. An increase of density would also favor faster evolution of interstellar dust through grain-grain collision. The low abundance of small dust grains in the densest parts of the front favors a coagulation process \citep[sticking of smaller grains onto bigger grains,][]{kohler2012}. 

The fact that the CMF and fragments of molecular clouds mass spectrum have a similar power-law index has been used as an argument in favor of turbulence being responsible for the fragmentation and structuration of matter in dense environments. On the other hand, because of the fact that most of these objects are gravitationally bound, it is difficult to exclude the role of gravity.
In this context, Draco is interesting as it reveals a clear example of dense structures that are formed through hydrodynamical processes combined with the thermal instability of \hi\  with the exact same statistical properties as gravitationally bound systems.\\

The small-scale structures of Draco have statistical properties of a thermally bistable turbulent flow. Numerical simulations \citep[e.g.,][]{saury2014} have shown that the formation of CNM structures happens when the WNM is thermally unstable (i.e., compressed) and if the cooling time is shorter than then dynamical time. These two conditions must be satisfied in order for the transition to occur. 
The cloud collision provides the necessary increase in WNM density but at the same time, the input of kinetic energy increases the amplitude of the turbulent motions, lowering the dynamical time. The formation of CNM can then only occur if part of the turbulent energy is dissipated such that dynamical time becomes larger than the cooling time. 
The small-scale structures of Draco have sizes slightly larger (0.1 to 0.5\,pc) than the WNM dissipation scale estimated here ($l_{\rm d} = 0.1$\,pc). This correspondance could be a coincidence; the minimum size of the structures found here is limited by the angular resolution and noise level of the data. Nevertheless, in the absence of higher resolution data, the observations presented here are compatible with the idea that the dissipation of turbulent energy of the WNM via the ambipolar diffusion provides favorable conditions for the formation of dense structures at scales close to $l_{\rm d}$.

\section{Conclusion}

\label{sec:conclusion}

   We presented \Herschel-SPIRE observations of the Draco nebula, a diffuse high Galactic latitude interstellar cloud located about $370\: pc$ above the Galactic plane. Draco is likely to be the result of the compression of diffuse gas of the outer WNM layer by the collision of a cloud falling from the Galactic halo. It offers a unique opportunity to study the formation of dense structures in a colliding flow.

The \Herschel\ data reveal the fine details of the structure of this cloud, especially of a front that shows a typical Rayleigh-Taylor instability structure. We were able to estimate the gas kinematic viscosity from the typical length of this structure. Combined with 21\,cm GBT data, we estimated the Reynolds number ($Re = 1600$), the energy transfer rate ($\epsilon=3.9 \times 10^{-3}$\,$L_\odot$\,$M_\odot^{-1}$), and the dissipation scale of the turbulent cascade ($l_d = 0.1$\,pc). The viscosity and dissipation scale are typical of values that would be found in the WNM in the case in which the turbulent energy dissipation is dominated by ambipolar diffusion. 

The SPIRE 250\,$\mu$m map reveals high contrast structures with a wealth of small-scale clumps. Using the {\em clumpfind} algorithm, a total of 5028 structures were identified, with physical sizes ranging from 0.08 to 0.5\,pc, where the lower limit is set by the angular resolution of the data. The mass spectrum is very close to that of prestellar cores (in terms of shape and range), even though the densities in Draco are 100 times lower. The high mass part of the mass spectrum is well represented by a power-law $dN/d\log(M) \propto M^{-1.4}$, such as that seen for prestellar cores and molecular clouds fragments. On the other hand, the majority of the structures (85\% in mass) are not gravitationally bound. This is corroborated by the fact that the density spectrum clearly follows a log-normal distribution. 

The results presented here reveal dense structures formed in a colliding flow where gravity is not a dominant process. We showed that these structures are likely to be the result of the transition from the WNM to CNM through the \hi\ thermal instability. We propose that the formation of dense structures is favored by the dissipation of the turbulent energy in the WNM through ambipolar diffusion, at a scale of $l_{\rm AD}=0.1$\,pc. The energy dissipation favors the formation of small CNM structures at scales slightly larger than $l_{\rm AD}$. The fact that these structures share many statistical properties with denser environments is an indication of the importance of interstellar turbulence and thermal instability in shaping the structure of the ISM.

\begin{acknowledgements}
   Herschel-SPIRE has been developed by a consortium of institutes led by Cardiff University (UK) and including Univ. Lethbridge (Canada); NAOC (China); CEA, LAM (France); IFSI, Univ. Padua (Italy); IAC (Spain); Stockholm Observatory (Sweden); Imperial College London, RAL, UCL-MSSL, UKATC, Univ. Sussex (UK); and Caltech, JPL, NHSC, Univ. Colorado (USA). This development has been supported by national funding agencies: CSA (Canada); NAOC (China); CEA, CNES, CNRS (France); ASI (Italy); MCINN (Spain); SNSB (Sweden); STFC, UKSA (UK); and NASA (USA).
\end{acknowledgements}

\bibliography{Article_v11}
\bibliographystyle{aa}

\begin{appendix}

\section{Estimating gas column density from dust emission}

\begin{figure*}[]
  \centering
  \includegraphics[draft=false, width=0.48\linewidth]{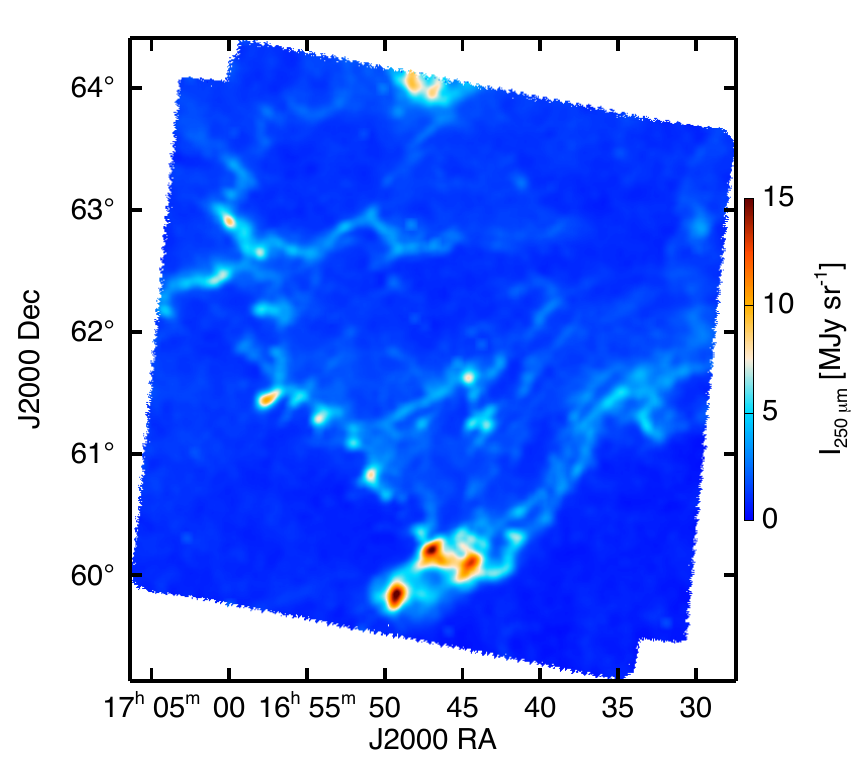}
  \includegraphics[draft=false, width=0.48\linewidth]{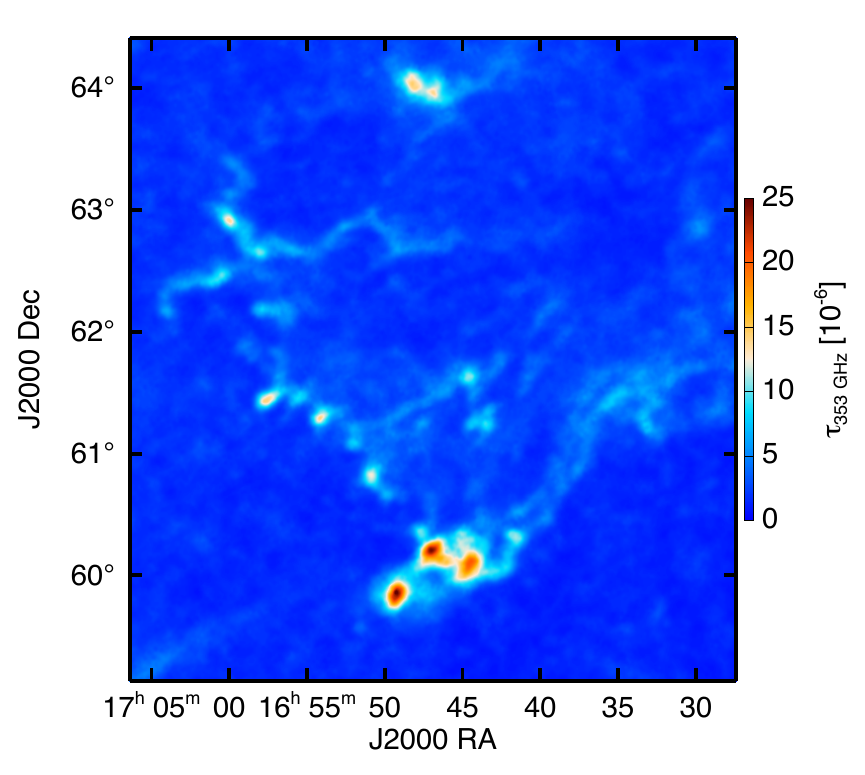}
  \includegraphics[draft=false, width=0.48\linewidth]{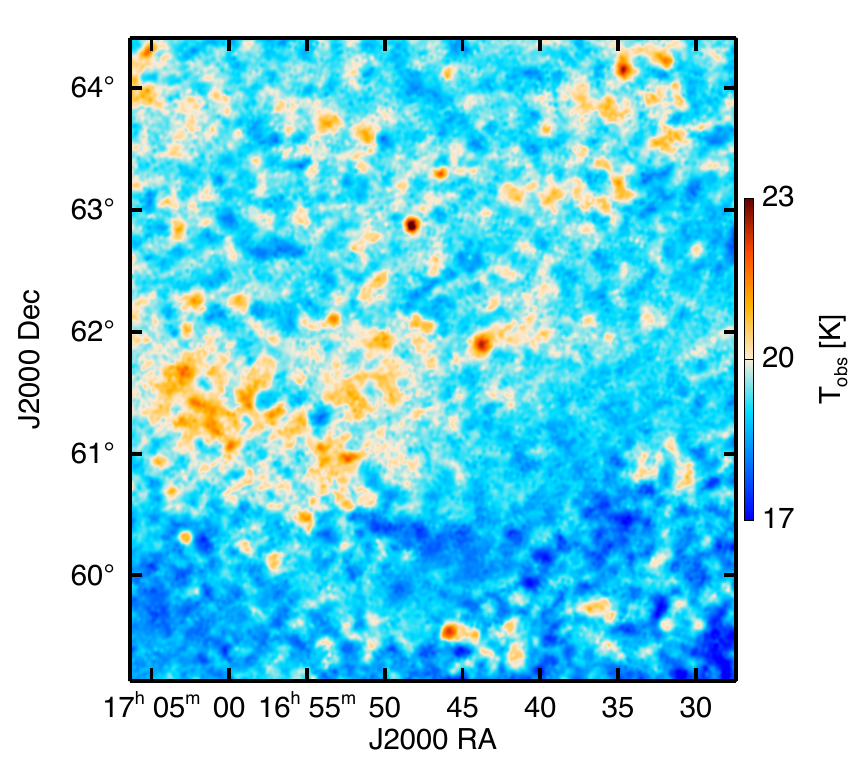}
  \includegraphics[draft=false, width=0.48\linewidth]{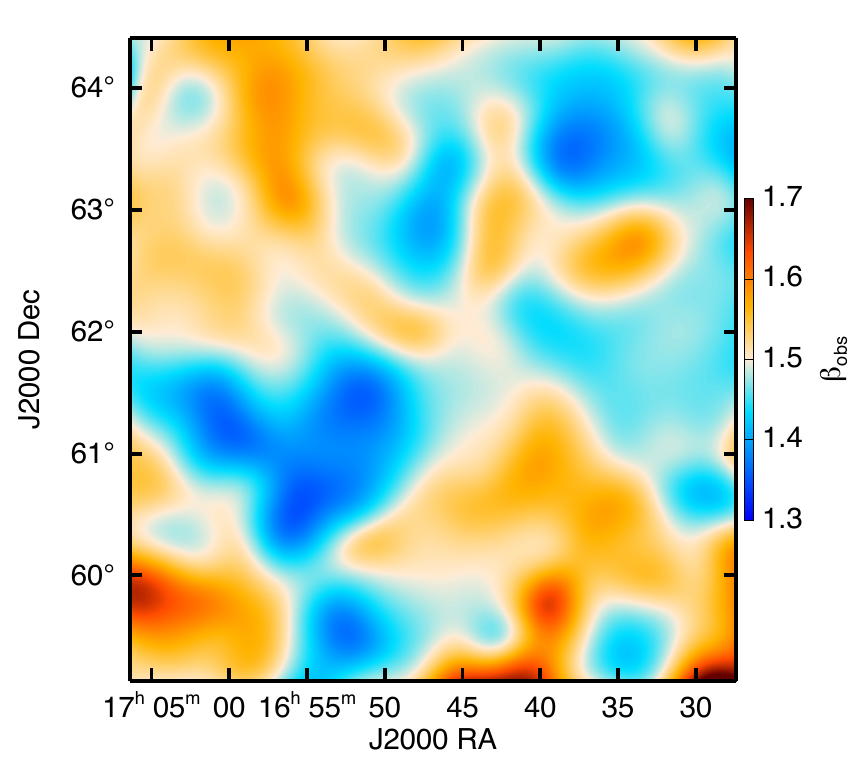}
  \caption{\label{fig:tau_T_beta_planck} Comparison of \Herschel-SPIRE and \Planck\ data of Draco. \Herschel-SPIRE at 250\,$\mu$m (top left), \Planck\ $\tau_{353}$ (top right), \Planck\ $T_{\rm obs}$ (bottom left), and \Planck\ $\beta_{\rm obs}$ (bottom right). All maps have an angular resolution of 5' except $\beta_{\rm obs}$ that is at 30'.}
\end{figure*}

\label{sec:convert_to_NH}

We are interested in using the large dust grain emission as a tracer of column density. To do so, it is customary to fit the spectral energy distribution of dust using a model to separate the effects of dust properties, dust temperature, and column density. The dust emission in the SPIRE wavelength range is dominated by big grains at thermal equilibrium with the ambient radiation field. It is often modeled as a modified blackbody spectrum,
i.e.,\begin{equation} \label{eq:modif-BB}
  I_\nu= \tau_{\nu_0}\, B_\nu(T_{\rm obs}) \, \cbra{\frac{\nu}{\nu_0}}^{\beta_{\rm obs}}
,\end{equation}
where $\tau_{\nu 0}$ is the optical depth at a reference frequency $\nu_0$, $T_{\rm obs}$ represents the dust equilibrium temperature, and $\beta_{\rm obs}$ is the dust spectral index.
For a constant dust emissivity and a constant dust-to-gas ratio, $\tau_{\nu_0}$ is proportional to gas column density. It can thus be used as a tracer of the structure of matter projected on the sky. To obtain $\tau_{\nu_0}$ it is common practice to fit the dust emission pixel by pixel, by combining PACS and SPIRE data for instance. This is unfortunately impossible in our case because of the failure of the PACS observations. 

A similar fit has been performed over the whole sky by \citet{planck_collaboration2014h} using \IRAS\ 100\,$\mu$m data and \Planck\ 350, 550, and 850\,$\mu$m data. The maps of $\tau_{\rm 353\,GHz}$, $T_{\rm obs}$, and $\beta_{\rm obs}$ for the Draco field are shown in Fig.~\ref{fig:tau_T_beta_planck}. These maps have an angular resolution of 5', 5', and 30', respectively. 
As a comparison, the SPIRE 250\,$\mu$m map convolved at 5' is also shown in Fig.~\ref{fig:tau_T_beta_planck}. The morphological resemblance of $I_{250\, \mu m}$ and $\tau_{\rm 353\, GHz}$ is striking, indicating that variations of dust properties and dust temperature do not dominate the observed emission fluctuations in the SPIRE bands. This is confirmed by looking at the maps of $T_{\rm obs}$ and $\beta_{\rm obs}$ computed by \citet{planck_collaboration2014h}. The $1 \sigma$ variations of $T_{\rm obs}$ and $\beta_{\rm obs}$ are only 3\% and the structure of Draco is barely visible in these parameter maps. This is in accordance with the fact that there are no local sources of heating photons and a relatively low extinction; locally the radiation field appears to be very uniform.

In addition we note the presence of small-scale structures in $T_{\rm obs}$, apparently unrelated to variations of the dust equilibrium temperature. These fluctuations were observed and quantified by \citet{planck_collaboration2014h}; they are caused by the cosmic infrared background anisotropies (CIBA). In such a diffuse high Galactic latitude field, the CIBA appear very strongly in the map of $T_{\rm obs}$ (see \citet{planck_collaboration2014h} for details).

\begin{figure}[]
  \centering
  \includegraphics[draft=false, width=\linewidth]{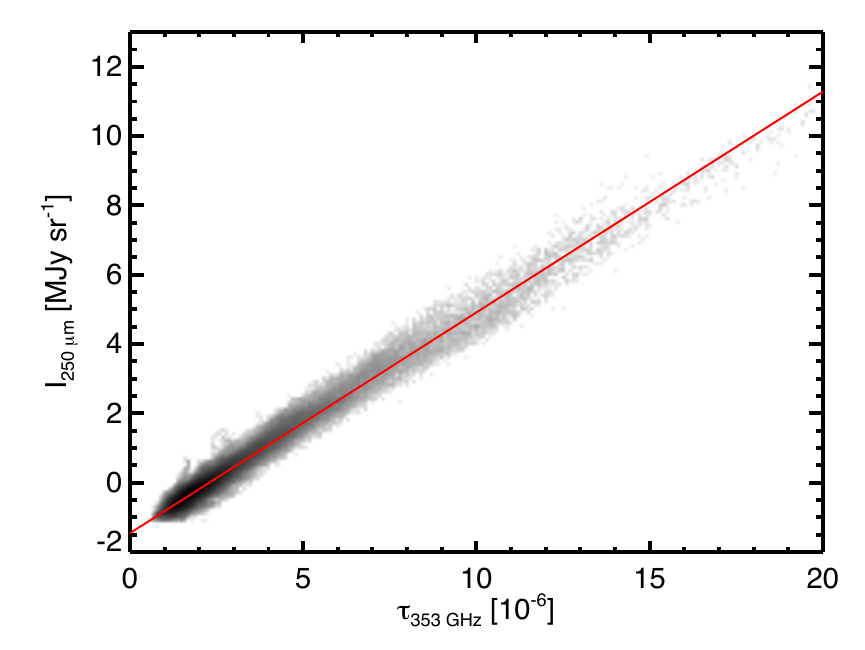}
  \caption{\label{fig:tau_vs_I250} \Herschel-SPIRE 250\,$\mu$m intensity, smoothed to 5', compared to \Planck\ optical depth at 353\,GHz. The red line is the linear relation $I_{250\,\mu m} = 6.37 \times 10^5 \, \tau_{\rm 353 \, GHz} - 1.46$. The intercept of this relation was removed from the $I_{\rm 250\, \mu m}$ map to set its zero level.}
\end{figure}

Even though the \Planck\ dust products are at significantly lower resolution than the SPIRE data, one could envisage estimating the dust column density by correcting for $T_{\rm obs}$ and $\beta_{\rm obs}$ obtained with \Planck. We conclude that the variations in $T_{\rm obs}$ that would be of interstellar origin are not strong enough to overcome the drawback of affecting the data with the CIBA. In addition the correlation of $I_{250 \mu m}$, convolved at 5', with $\tau_{\rm 353 GHz}$ is extremely tight (see Fig.~\ref{fig:tau_vs_I250}), with residuals compatible with CIBA. For those reasons we decided to simply multiply $I_{250\, \mu m}$ by a single factor to translate it to hydrogen column density, $N_H$. 
To do so, we use the $I_{250\, \mu m}$-$\tau_{\rm 353\, GHz}$ correlation shown in Fig.~\ref{fig:tau_vs_I250} to set the zero level of the SPIRE map. Then, combined with the relationship $\tau_{\rm 353\, GHz} = 6.3 \times 10^{-27} \, N_{HI}$ that \citet{planck_collaboration2014h} found by comparing \Planck\ and 21\,cm data at high Galactic latitude, we converted the 250\,$\mu$m emission to hydrogen column density using the following relation:
\begin{equation}
N_{\rm H} = 2.49\times 10^{20} \, I_{\rm 250\, \mu m}
  ,\end{equation}
where $I_{\rm 250\, \mu m}$ is the zero level corrected map, expressed in MJy\,sr$^{-1}$, and $N_{\rm H}$ is the total hydrogen column density (\hi\ and H$_2$) expressed in cm$^{-2}$.
Because we assumed constant values for $T_{\rm obs}$ and $\beta_{\rm obs}$, no color correction is needed in the correspondence between $I_{\rm 250\, \mu m}$, $\tau_{\rm 353\, GHz}$, and $N_{H}$. 

\section{Size of structures}

\label{sec:size}

Each structure is composed of a number of pixels in the map,where  each pixel $i$ is defined by a position [$X_i$, $Y_i$] and a column density $N_{{\rm H}, i}$. The structures are not assumed to be round; we estimate the length of their short and long axis using the inertia matrix \citep{hennebelle2007a,saury2014}
      \begin{equation}
M = \begin{bmatrix}
       \sigma^2(X) & \sigma^2(XY)          \\
       \sigma^2(XY) & \sigma^2(Y)          \\
     \end{bmatrix}
    ,\end{equation}
      where      
      \begin{equation}
        \sigma^2(X) = \frac{1}{N_{\rm H tot}} \, \sum_i N_{{\rm H}, i} \, ( X_i - \langle X \rangle )^2
      \end{equation}
\begin{equation}
      \sigma^2(Y) = \frac{1}{N_{\rm H tot}} \, \sum_i N_{{\rm H}, i} \, (Y_i - \langle Y \rangle)^2
\end{equation}
\begin{equation}
      \sigma^2(XY) = \frac{1}{N_{\rm H tot}} \, \sum_i N_{{\rm H}, i} \, (X_i - \langle X \rangle) \, (Y_i - \langle Y \rangle)
\end{equation}
with
\begin{equation}
      N_{\rm H tot} = \sum_i N_{{\rm H}, i}
\end{equation}
\begin{equation}
      \langle X \rangle = \frac{1}{N_{\rm H tot}} \, \sum_i N_{{\rm H}, i} \, X_i
\end{equation}
\begin{equation}
    \langle Y \rangle = \frac{1}{N_{\rm H tot}} \, \sum_i N_{{\rm H}, i} \, Y_i.
\end{equation}

    The two eigenvalues of $M$ correspond to the smallest and largest semi-axes of the structure,
    denoted $L_{\rm min}$ and $L_{\rm max}$,  respectively. From these, we define a typical size for the structure, i.e.,
    \begin{equation}
    L = \left( L_{\rm max} \, L_{\rm min}^2 \right)^{1/3}.
\end{equation}
Here we make the hypothesis that an elongated structure projected on the sky is more likely to have a depth along the line of sight that corresponds to the smallest projected size on the sky.

We check that for a uniformly weighted circle of radius $R$, this method gives $L=R$. More generally, $L_{\rm min}$
and $L_{\rm max}$ correspond to the half length of an ellipse, whatever its orientation with respect to the $X-Y$ plane. For a point source with a Gaussian distribution of brightness of width $\sigma$,
the typical size is $L = 1.77 \, \sigma$ or $L = 0.75\, FWHM$.

We also consider another way of defining the size of a cloud, often used in the literature, where $R=\sqrt{A/\pi}$, with $A$ being the surface of the structure (in pc$^2$) proportional to the number of pixels of the structure, $N_{\rm pix}$. These two methods give very concordant results indicating that structures are not very elongated.

  \end{appendix}

\end{document}